\documentclass[12pt,preprint]{aastex}
\RequirePackage{epstopdf}

\slugcomment{Draft of \today}

\shorttitle{CNR Paper}
\shortauthors{Lau et al.}

\usepackage{graphicx}
\usepackage[section] {placeins}

\newcommand{\beq}{\begin{equation}}
\newcommand{\eeq}{\end{equation}}

\begin{document}

\title{SOFIA/FORCAST Imaging of the Circumnuclear Ring at the Galactic Center}

\author{R. M. Lau\altaffilmark{1},
T. L. Herter\altaffilmark{1},
M. R. Morris\altaffilmark{2},
E. E. Becklin\altaffilmark{2,3},
J. D. Adams\altaffilmark{1}
}

\altaffiltext{1}{Astronomy Department, 202 Space Sciences Building, Cornell University, Ithaca, NY 14853-6801, USA}
\altaffiltext{2}{Department of Physics and Astronomy, University of California, Los Angeles, 430 Portola Plaza, Los Angeles, CA 90095-1547, USA}
\altaffiltext{3}{SOFIA Science Center, Universities Space Research Association, NASA Ames Research Center, MS 232, Moffett Field, CA 94035, USA}

\begin{abstract}
We present $19.7$, $31.5$, and $37.1$ $\mu$m images of the inner 6 pc of the Galactic Center of the Milky Way with a spatial resolution of 3.2 - 4.6'' taken by the Faint Object Infrared Camera on the Stratospheric Observatory for Infrared Astronomy (SOFIA). The images reveal in detail the ``clumpy" structure of the Circumnuclear Ring (CNR)--the inner-edge of the molecular torus orbiting the supermassive black hole at the Galactic Center--and the prominent streamers of hot, ionized gas and dust within the CNR that compose the HII region Sgr A West. The CNR exhibits features of a classic HII region: the dust emission at 19.7 $\mu$m closely traces the ionized gas emission observed in the radio while the 31.5 and 37.1 $\mu$m emission traces the photo-dissociation region beyond the ionized gas. The 19.7/37.1 color temperature map reveals a radial temperature gradient across the CNR with temperatures ranging from 65-85 K, consistent with the prevailing paradigm in which the dust is centrally heated by the inner cluster of hot, young stars. We model the 37.1 $\mu$m intensity of the CNR as an inclined ($\theta_i=67^{\circ}$) ring with a thickness and radius of 0.34 pc and 1.4 pc, respectively, and find that it is consistent with the observed 37.1 $\mu$m map of the CNR. The 37.1 $\mu$m optical depth map also reveals the clumpy dust distribution of the CNR and implies a total gas mass of $\sim610\,M_\odot$. Dense (5 to 9 $\times 10^4\,\mathrm{cm}^{-3}$) clumps with a FWHM of $\sim0.15$ pc exist along the inner edge of the CNR and shadow the material deeper into the ring. We find that the clumps are unlikely to be long-lived structures since they are not dense enough to be stable against tidal shear from the supermassive black hole and will be sheared out on a timescale of an orbital period ($\sim10^5$ yrs).

\end{abstract}

\maketitle

\section{Introduction}

As in many gas-rich spiral galaxies, our Galaxy has an orbiting torus of warm, dense gas circling the central, supermassive black hole: the Circumnuclear Disk (CND) (Genzel et al. 1985). Due to heavy visual extinction by interstellar dust the Galactic Center (GC) has been studied intensely at infrared and radio wavelengths. Early infrared observations taken aboard the Kuiper Airborne Observatory (KAO) revealed the presence of strong, extended IR emission from the inner few pc of the GC interpreted as a ring of hot dust (Becklin, Gatley, \& Werner 1982). Radio observations revealed the compact radio source, Sgr A*, located near the center of thermal streamers of ionized gas referred to as the ``minispiral", or Sgr A West, as well as emission from the non-thermal shell, Sgr A East (Balick \& Brown 1974; Yusef-Zadeh \& Morris 1987; Roberts \& Goss 1993). Sgr A*, the dynamic center of the Galaxy ($d_\mathrm{GC}\sim8$ kpc (Reid et al. 1993)), is identified with a 4 million Solar mass (Ghez et al. 2008; Gillessen et al. 2009) supermassive black hole.

Surrounding Sgr A* and extending from 1.4 pc to beyond $\sim5$ pc is the clumpy torus of dense molecular gas and dust known as the Circumnuclear Disk. Atomic fine-structure line observations (63 $\mu$m [O I]) of the Sgr A West region revealed that the CND indeed circulates around Sgr A* (Genzel et al. 1984; Jackson et al. 1993). Much of the ionized gas features observed at Sgr A West are coincident with the inner edge of the CND, indicating that these features are HII regions at the inward-facing edge of the CND excited by Lyman continuum radiation from a central source; radial velocity measurements of hydrogen radio recombination lines verified that the ionized features along the Western Arc can be modeled in terms of a circular orbit around Sgr A* (Roberts \& Goss 1993). Mid to far-IR observations of the warm dust emission from Sgr A West revealed dust temperatures consistent with heating from a central source (Davidson et al 1992; Telesco et al. 1996). With the Kuiper Widefield Infrared Camera (KWIC) (Stacey et al. 1993) on the KAO Latvakoski et al. (1999) captured the full ring of illuminated dust in the far-IR at the inner-edge of the CND, including the faint eastern inner-edge unobserved in the radio due to shadowing from the in-falling streamers. The inconsistent appearance of the CND traced by different wavelengths arising from different physical regions has prompted us to refer to the illuminated inner edge of the CND as the ``Circumnuclear Ring'' (CNR). Thus, in our framework, the CNR is the ring-like structure defined by the emission from the ionized gas and hot dust at the inner edge of the CND, whereas the CND is characterized by its disk-like appearance from molecular observations.  

Several different interpretations of the CNR morphology exist. Zhao et al. (2009) suggest from the radial velocity observations of the H92$\alpha$ recombination line that the Western Arc of the CNR as well as the Northern Arm and East-West Bar--the two streamers plunging towards Sgr A* (Lacy et al. 1980; Serabyn \& Lacy 1988)--can be modeled as Keplerian orbits around Sgr A*.  They find that the Western Arc exhibits a near-circular orbit, which is consistent with the morphological interpretation suggested by Latvakoski et al. (1999) from observations of hot dust emission. Latvakoski et al. (1999) treat the Northern Arm and East-West Bar as parabolic streamers with Sgr A* at the focus and having inclinations out of the plane of the CNR. Lacy et al. (1991) and Irons et al. (2012) depart from the standard circular ring interpretation of the CNR and instead present a unified model of the Northern arm and Western Arc of the CNR as a one-armed spiral density wave using velocity data from [NeII] emission.

The morphology and physical properties of the CNR (and CND) largely tie in to their interpretation as either transient or long-lived structures. Etxaluze et al. (2011) fit spectral energy distributions to their observed far-IR dust emission of the CND to determine a mass of 5 $\times\,10^4\,\mathrm{M}_\odot$, which they argue is largely dominated by a distribution of cold dust ($\sim20$ K). The warmer dust ($\sim80$ K) in the CND defines the CNR; it is traced by the emission at 31 and 37 $\mu$m and implies a total mass of $\sim1000\,\mathrm{M}_\odot$ (Latvakoski et al. 1999). The total CNR mass is consistent between the two separate IR observations performed by Etxaluze et al. (2011) and Latvakoski et al. (1999); Etxaluze et al. (2011) determine a larger total CND gas mass because they observe at longer wavelengths and are thereby sensitive to the larger quantity of cooler dust unobserved by Latvakoski et al. (1999). Recent SOFIA work by Requena-Torres et al. (2012) study the CO rotational line transitions in the CND that imply a total CND mass of $\sim10^4\,\mathrm{M}_\odot$, which is comparable to that of Etxaluze et al. (2011). Observations of different molecular tracers such as HCN (Christopher et al. 2005; Montero-Casta\~no et al. 2009) led to a much larger estimate for the total CND mass on the order of $10^6\,\mathrm{M}_\odot$ under the assumption that the observed clumps of molecular material in the CND are self-gravitating.

In this paper we present 19.7, 31.5, and 37.1 $\mu$m observations tracing the hot dust emission from the inner 6 pc of the Galactic Center taken by FORCAST aboard the Stratospheric Observatory for Infrared Astronomy (SOFIA). Our images reveal the CNR and ``minispiral" at the highest spatial resolution in the mid-IR so far and allow us to study the morphology and structure of the CNR dust emission in great detail. We determine the morphological properties of the CNR and generate a dust emission ring model to compare against our observations. With our multi-wavelength coverage of the CNR we study the physical properties of the emitting dust and derive constraints to the total mass and density. At the inner edge of the CNR we resolve small ($\sim0.15$ pc), density-enhanced clumps that do not appear dense enough to survive the tidal shear and are therefore unlikely to be permanent features in the ring.

\section{Observations and Data Reduction}

Observations were made using FORCAST (Herter et al. 2012) on the 2.5 m telescope aboard SOFIA. FORCAST is a  $256 \times 256$ pixel dual-channel, wide-field mid-infrared camera sensitive from $5 - 40~\mu\mathrm{m}$ with a plate scale of $0.768''$ per pixel and field of view of $3.4'\,\times\,3.2'$.
The two channels consist of a short wavelength camera (SWC) operating at $5 - 25~\mu\mathrm{m}$ and a long wavelength
camera (LWC) operating at $28 - 40~\mu\mathrm{m}$. An internal dichroic beam-splitter enables simultaneous observation from both long and short wavelength cameras. A series of bandpass filters is used to image at selected wavelengths.

SOFIA/FORCAST observed the Circumnuclear Disk during Basic Science flights 63 and 64 on June 4, 2011 (altitude $\sim$ 41,000 ft.)
and June 8, 2011 (altitude $\sim$ 43,000 ft.), respectively, at wavelengths of 19.7, 31.4, and 37.1 $\mu\mathrm{m}$.
The wavelengths 19.7 and 31.4 $\mu$m were observed simultaneously in dual channel mode, while
the 37.1 $\mu$m observations were made in single channel mode. Chopping and nodding were used to remove  the sky and telescope thermal backgrounds. An asymmetric chop pattern was used to place the source on the telescope
axis which eliminates optical aberrations (coma) on the source. The chop throw was $7'$ at a frequency of $\sim 4$ Hz.
The off-source chop fields (regions of low mid-infrared Galactic emission) were selected from KAO/KWIC images of the Galactic
Center at $30-38~\mu$m (Latvakoski et al. 1999) and a combined Multiband Imaging Photometer for Spitzer (MIPS) $24~\mu$m and Midcourse Space Experiment (MSX) $21~\mu$m image of the Galactic Center that was kindly provided to us by F. Yusef-Zadeh (Yusef-Zadeh, priv. comm.).
The source was dithered over the focal plane to allow removal of bad pixels and mitigate response variations.
The integration time at each dither position was $\sim 30$ sec. The quality of the images was consistent with
near-diffraction-limited imaging at $19.7 - 37.1~\mu$m; the full width at half maximum (FWHM) of the point spread function (PSF) was
$3.2''$ at $19.7~\mu$m and $4.6''$ at 37.1 $\mu$m.

The acquired 30 sec integrated images were reduced and combined at each wavelength according to the pipeline steps described
in Herter et al. (2013). The calibration factors that were applied to the data numbers were the average calibration factors derived from calibration observations taken over the Early SOFIA Science phase, adjusted to those of a flat spectrum ($\nu F_\nu = \rm{constant}$) source. The $3\sigma$
uncertainty in the calibration factors is $\pm20\%$.

The relatively high brightness and contrast of the CNR region on the FORCAST detectors triggered cross-talk artifacts across the images and produced some suppression of signal (i.e., a``bowl'') over regions of the detector. The ``bowl" is much less prominent in the images taken during flight 63 where the readout frame-rate was higher than in flight 64. We have removed much of the cross-talk artifacts using the channel subtraction algorithm, however, a residual remains along the rows that experienced the largest magnitude of the cross-talk signal. In addition to the ``bowl,'' the images taken during flight 64 have a compact negative flux region north of the CNR due to a bright source present in the "off'' position of the chop. The total absolute flux  of the ``bowl" and the ``negative source" is  $\sim10\%$ of observed flux from the CNR.

Two corrections were applied to the images to address the ``negative source" and the ``bowl." The images from flight 63 (where the ``negative source" did not appear) were used to acquire a background flux level in the ``negative source" region for flight 64. This was accomplished by subtracting the flight 64 images form the flight 63 images to characterize the flux of the ``negative source" region and then remove it from the flight 64 images. By choosing regions off the source and interpolating across the source we were also able to remove the ``bowl'' at the penalty of removing any larger scale extended emission.

\section{Results and Analysis}

\begin{figure}[h]
	\centerline{\includegraphics[scale=0.5]{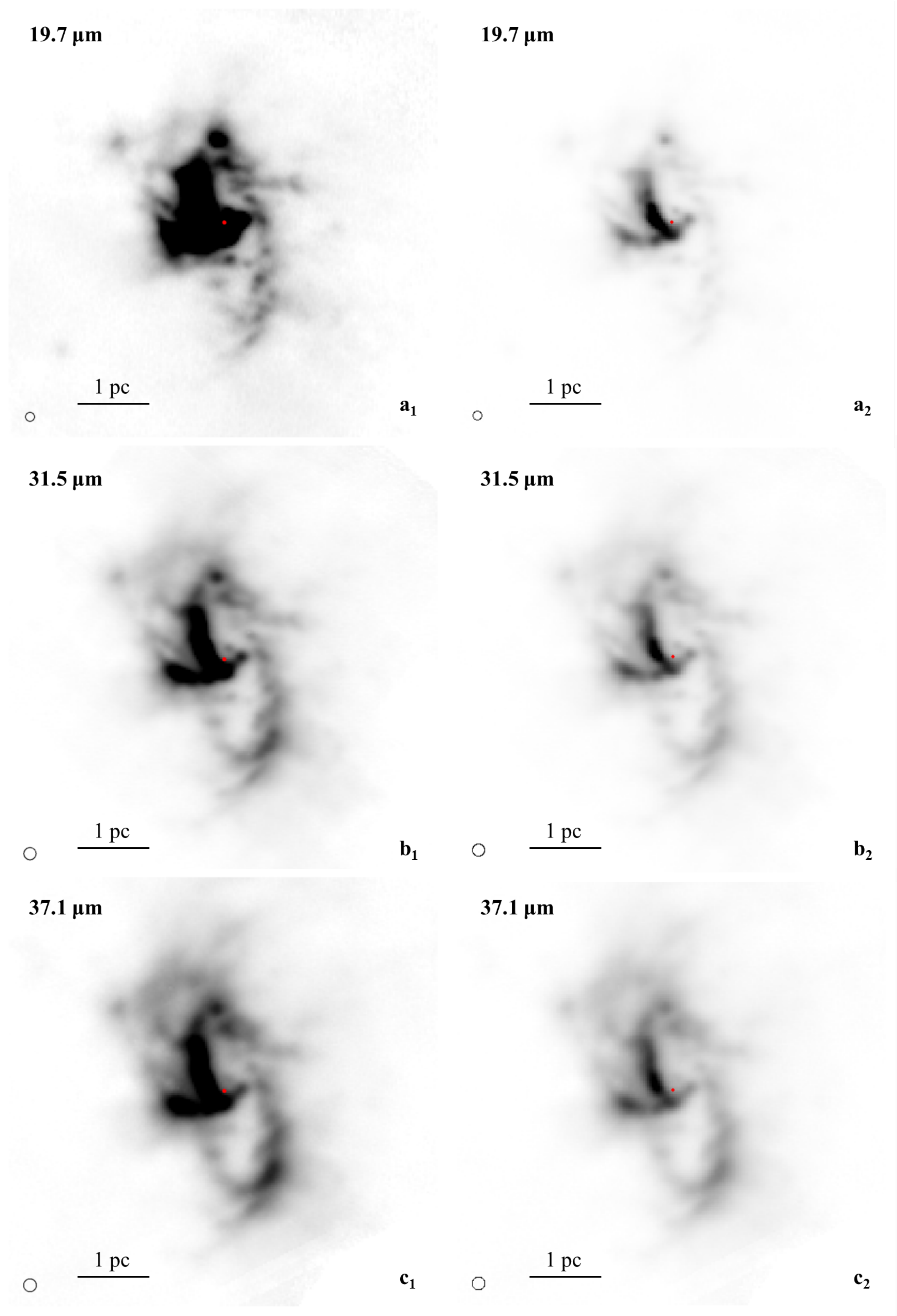}}
	\caption{Observed 19.7 (a), 31.5 (b), and 37.1 (c) $\mu$m images of the inner 6 pc of the Galactic Center with the red dot indicating the location of Sgr A* ($17^\mathrm{h}45^\mathrm{m}40^\mathrm{s}.0409$, $-29^{\circ}00'28.''118$). The approximate beamsizes are shown in the lower left corner in each image. In column 1 the images are stretched linearly to show the emission from the ring. In column 2 the images are stretched linearly to show the emission from the Northern Arm and East-West Bar.}
	\label{fig:CNRdustem}
\end{figure}

\begin{figure}[h]
	\centerline{\includegraphics[scale=.5]{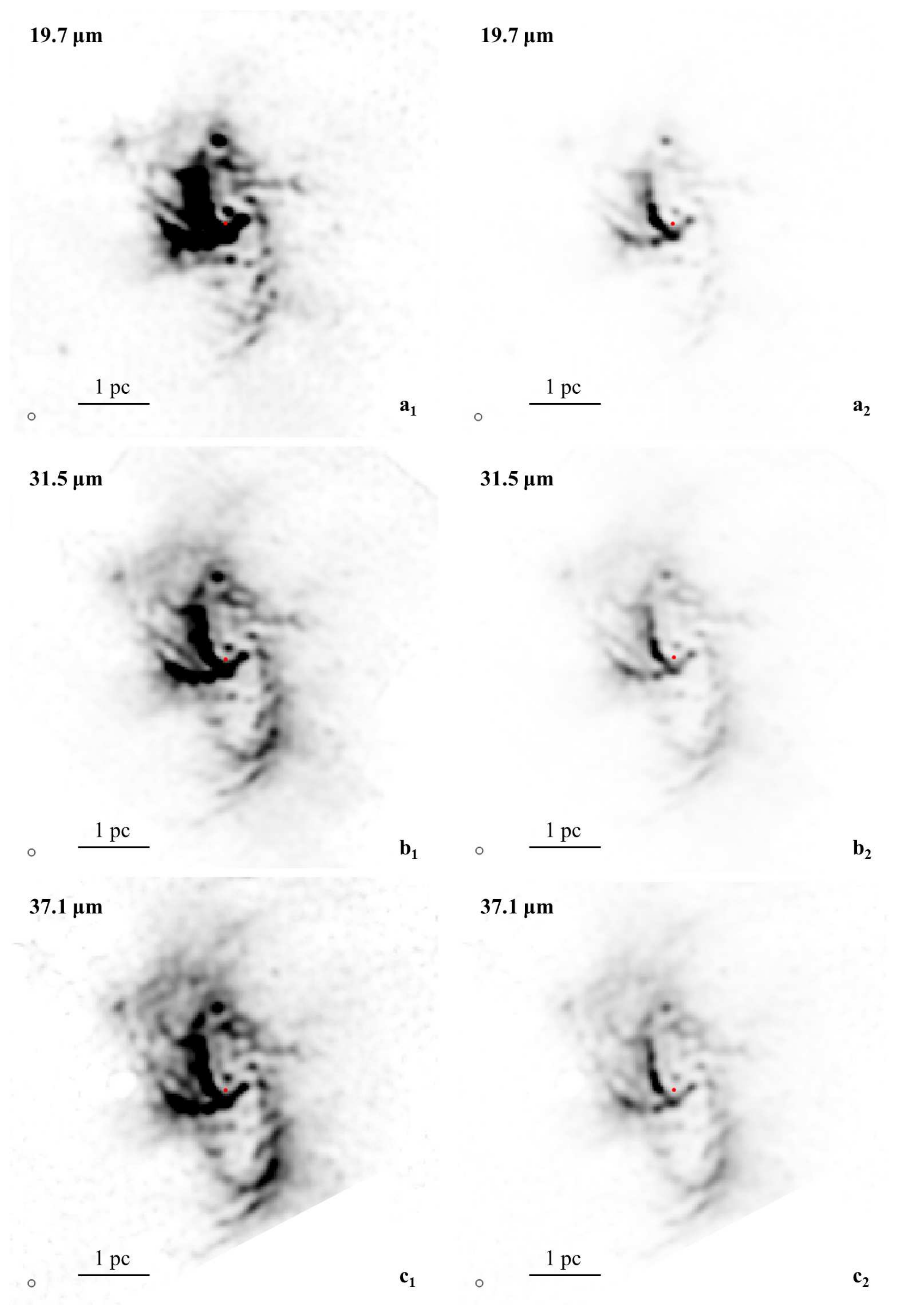}}
	\caption{Deconvolved 19.7 (a), 31.5 (b), and 37.1 (c) $\mu$m images of the inner 6 pc of the Galactic Center to a beamsize of 2.5'' (shown in the lower left). The red dot indicates the location of Sgr A* ($17^\mathrm{h}45^\mathrm{m}40^\mathrm{s}.0409$, $-29^{\circ}00'28.''118$). In column 1 the images are stretched linearly to show the emission from the ring. In column 2 the images are stretched linearly to show the emission from the Northern Arm and East-West Bar.}
	\label{fig:CNRdustemdecrec}
\end{figure}

\begin{figure}[h]
	\centerline{\includegraphics[scale=.5]{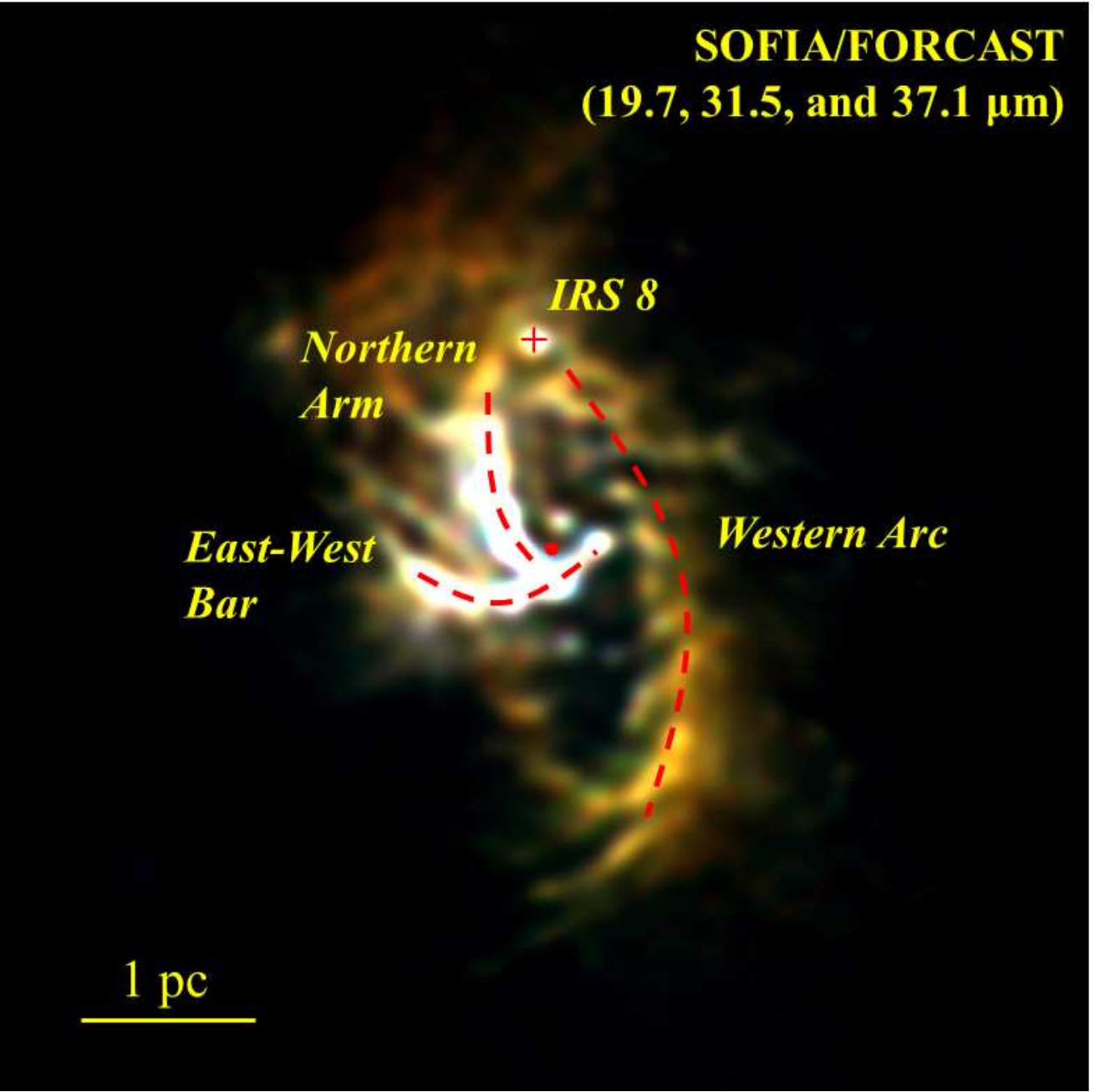}}
	\caption{Deconvolved (2.5'' beamsize) false color image of the CNR and Sgr A West made from the three FORCAST filters: 19.7 $\mu$m  - blue, 31.5 $\mu$m  - green, 37.1 $\mu$m - red. The components of the CNR and Sgr A West are traced by the dotted lines. The locations of Sgr A* and IRS 8 are indicated by the dot and cross, respectively.}
	\label{fig:CNRfc}
\end{figure}

The 19.7, 31.5, and 37.1 $\mu$m images shown in Fig.~\ref{fig:CNRdustem} reveal the warm dust emission of the Circumnuclear Ring (CNR) and Sgr A West in the inner 6 pc (2.6' $\times$ 2.6') of the Galactic Center. The J2000.0 coordinates of Sgr A* are $17^\mathrm{h}45^\mathrm{m}40^\mathrm{s}.0409$, $-29^{\circ}00'28.''118$ (Reid \& Brunthaler 2004). Since the spatial resolution is wavelength dependent we use Richardson-Lucy deconvolution routine to improve the spatial resolution and provide a uniform 2.5'' full-width at half maximum (FWHM) Gaussian PSF at all wavelengths. The deconvolved images have an effective resolution of 2.5'' and are shown in Fig.~\ref{fig:CNRdustemdecrec} along with a 19.7, 31.5, and 37.1 $\mu$m false color image in Fig.~\ref{fig:CNRfc}. We refer to the deconvolved images in our analyses for the rest of this paper.

The components of the Circumnuclear Ring (CNR) and Sgr A West are clearly evident in Fig.~\ref{fig:CNRdustemdecrec} and \ref{fig:CNRfc}. We also observe the prominent emission from the warm dust associated with the Northern Arm and East-West Bar, streamers of ionized gas falling inwards to Sgr A* (Serabyn \& Lacy 1985); however, in this paper we primarily focus on the properties of the CNR. Dust emission is observed from around the full CNR but is most prominent along the Western Arc; the flux from the eastern side of the CNR is $\sim1/3$ of that from the Western Arc. In the cavity to the south of Sgr A* and enclosed by the southern portion of the CNR the flux drops to $\sim10\%$ of the Western Arc flux. We observe several features around the ring extending into the southern cavity, some of which have ionized gas emission counterparts as seen in Paschen-$\alpha$ (Wang et al. 2010; Dong et al. 2011) and the radio continuum (Yusef-Zadeh \& Morris 1987). The northern region of the CNR is much more diffuse and undefined than the southern region where the southern edge of the Western Arc defines a sharp boundary to the inner cavity. We resolve the ``clumpiness'' of the CNR and observe small ``clumps" (FWHM $\sim 4-5''$) along the inner edge of the ring; these are discussed in greater detail in Sec. 4.5.

\subsection{Circumnuclear Ring Morphology}

\begin{figure}[h]
	\centerline{\includegraphics[scale=.5]{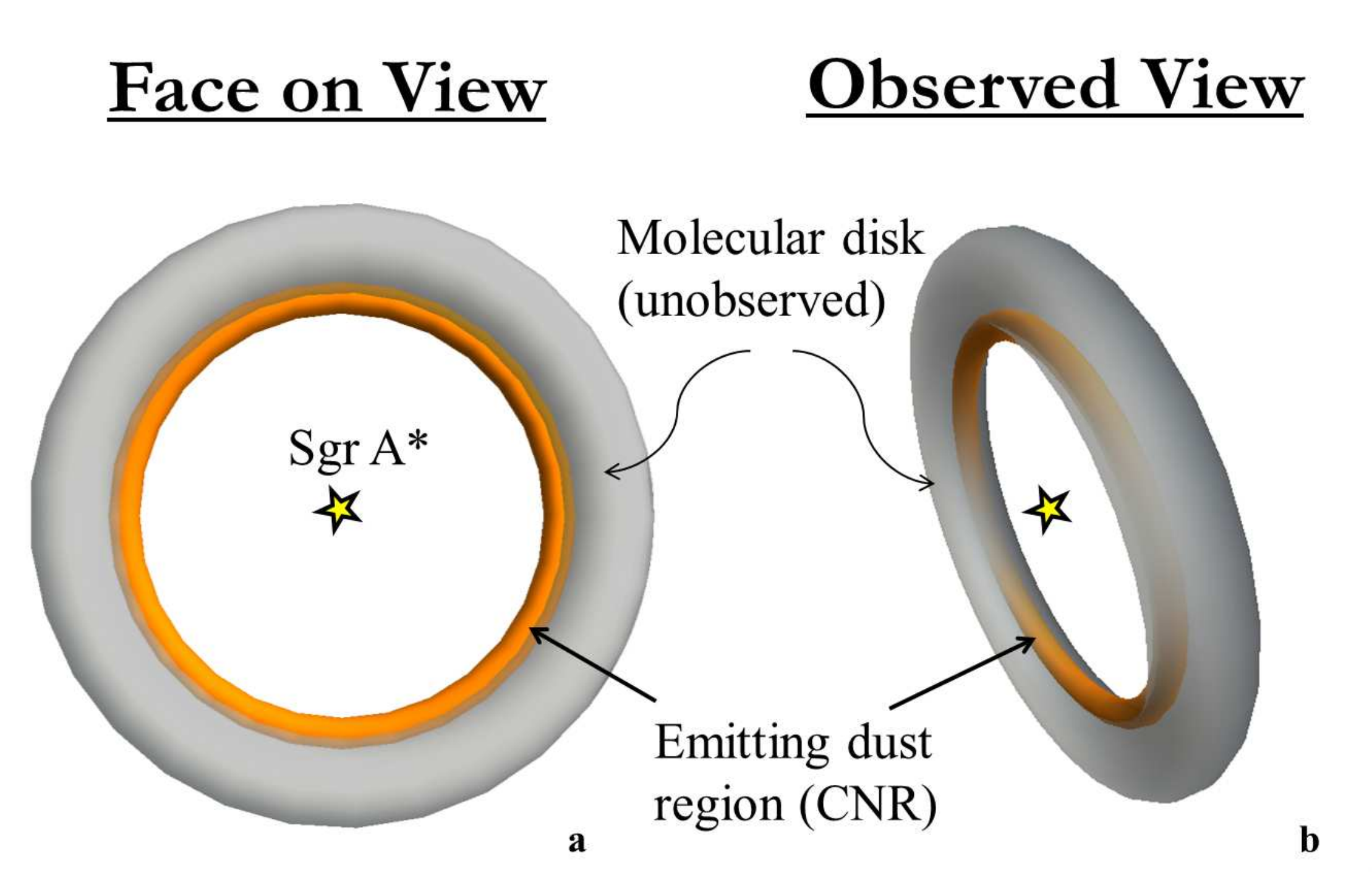}}
	\caption{Schematic diagram of the geometry of the Circumnuclear Ring (CNR). The diagram illustrates that the CNR is the illuminated inner edge (orange) of a larger disk of cool dust and gas (grey) unobserved at our wavelengths. The face on view model (a) shows the ring-like structure of the CNR centered on Sgr A*. The observed view model (b) is the same model as (a) inclined by $67^{\circ}$ with respect to the plane of the paper and rotated to align with the observed CNR in equatorial coordinates. The north and south regions of the CNR are brightened in (b) to reflect the effect of the inclination increasing the column density of the emitting dust along lines of sight at those regions.}
	\label{fig:CNRview}
\end{figure}

\begin{figure}[h]
	\centerline{\includegraphics[scale=.75]{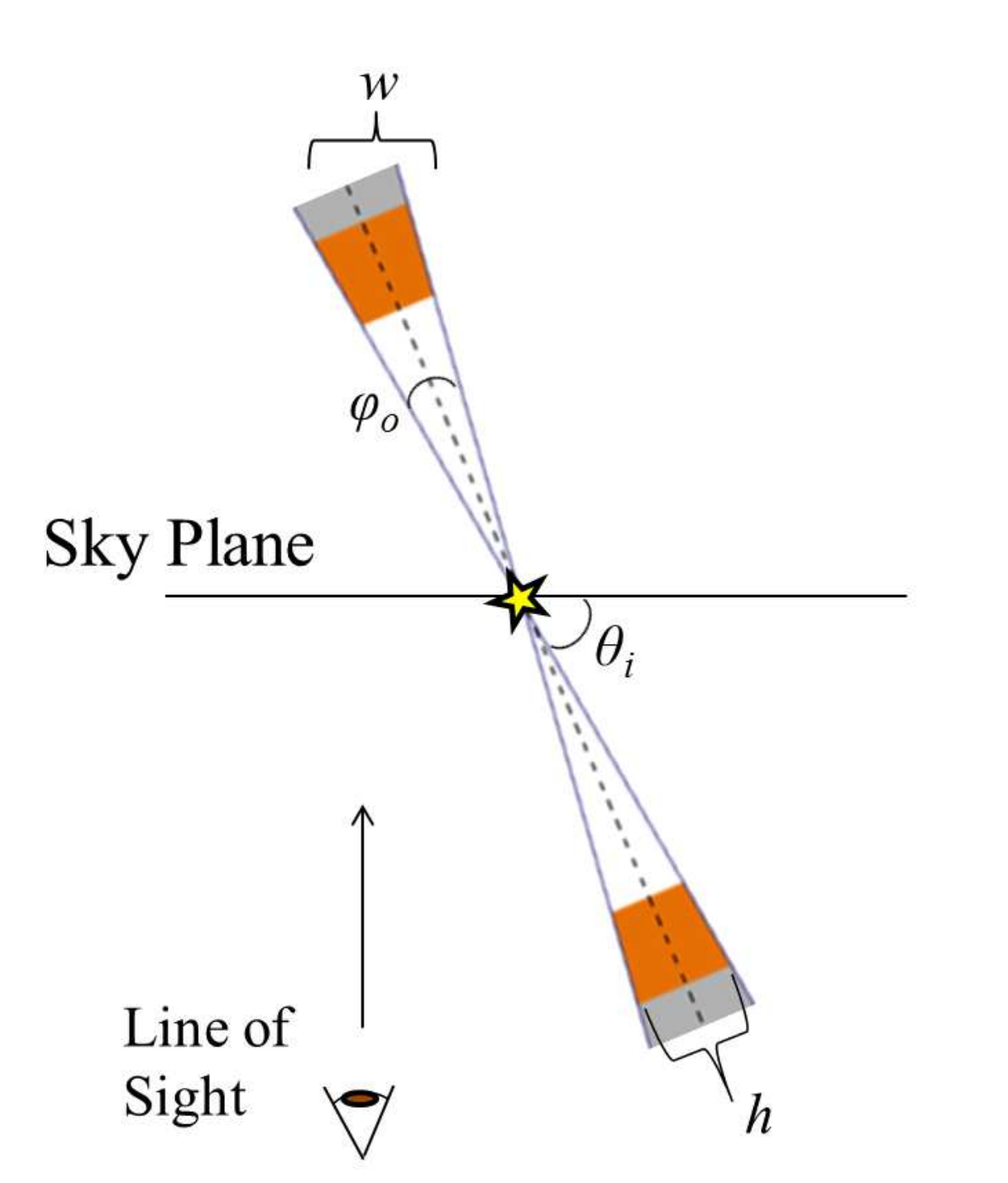}}
	\caption{Schematic cross section through the CNR (orange) and the unobserved, cool dust and gas disk (grey). The morphological parameters shown are the observed ring width $w$, the opening angle $\phi_0$, the inclination angle $\theta_i$, and the ring height $h$. }
	\label{fig:CNRCross}
\end{figure}

The deconvolved images (Fig.~\ref{fig:CNRdustemdecrec}) display the warm dust emission from the CNR in unprecedented detail. The schematic diagram shown in Fig.~\ref{fig:CNRview} illustrates that the CNR is observed as an ellipse due to the high inclination. Fig.~\ref{fig:CNRview} also conveys the distinct difference between the CNR and the larger, cooler distribution of gas and dust surrounding Sgr A* observed in molecular observations (G\"usten et al. 1987): the CNR is the illuminated inner edge of the cool disk of gas and dust, the Circumnuclear Disk, or CND. Assuming the CNR is a circular ring centered on Sgr A* we extract the properties of the projected CNR ellipse: the semi-minor axis is $14''\,\pm2''$ (0.54 pc $\pm 0.08$ pc), the semi-major axis is $36''\,\pm2''$ (1.4 pc $\pm 0.08$ pc), and the inclination is therefore $67^{\circ}\,\pm5^{\circ}$, which is consistent with the angle derived by Latvakoski et al. (1999). In addition to extracting the geometric properties of the projected CNR ellipse we determine the projected width of the ring, from which we can derive its height and opening angle (Fig.~\ref{fig:CNRCross}). The projected width of the ring is $\sim12''\pm1.7''$ (0.45 pc) along the minor axis and $\sim11''\pm1.7''$ (0.4 pc) along the major axis. Since the observed ring width along the major axis is the true, unprojected width of the ring we can use that and the measured minor axis ring width, assuming they are the same, to derive the opening angle and the height. We assume the geometry shown in Fig.~\ref{fig:CNRCross} and determine an opening angle of $\sim 12^{\circ}\pm 3^{\circ}$ and a ring height of $\sim0.29\pm0.08$ pc.

\subsection{CNR as a Classic HII Region}

\begin{figure}[h]
	\centerline{\includegraphics[scale=0.75]{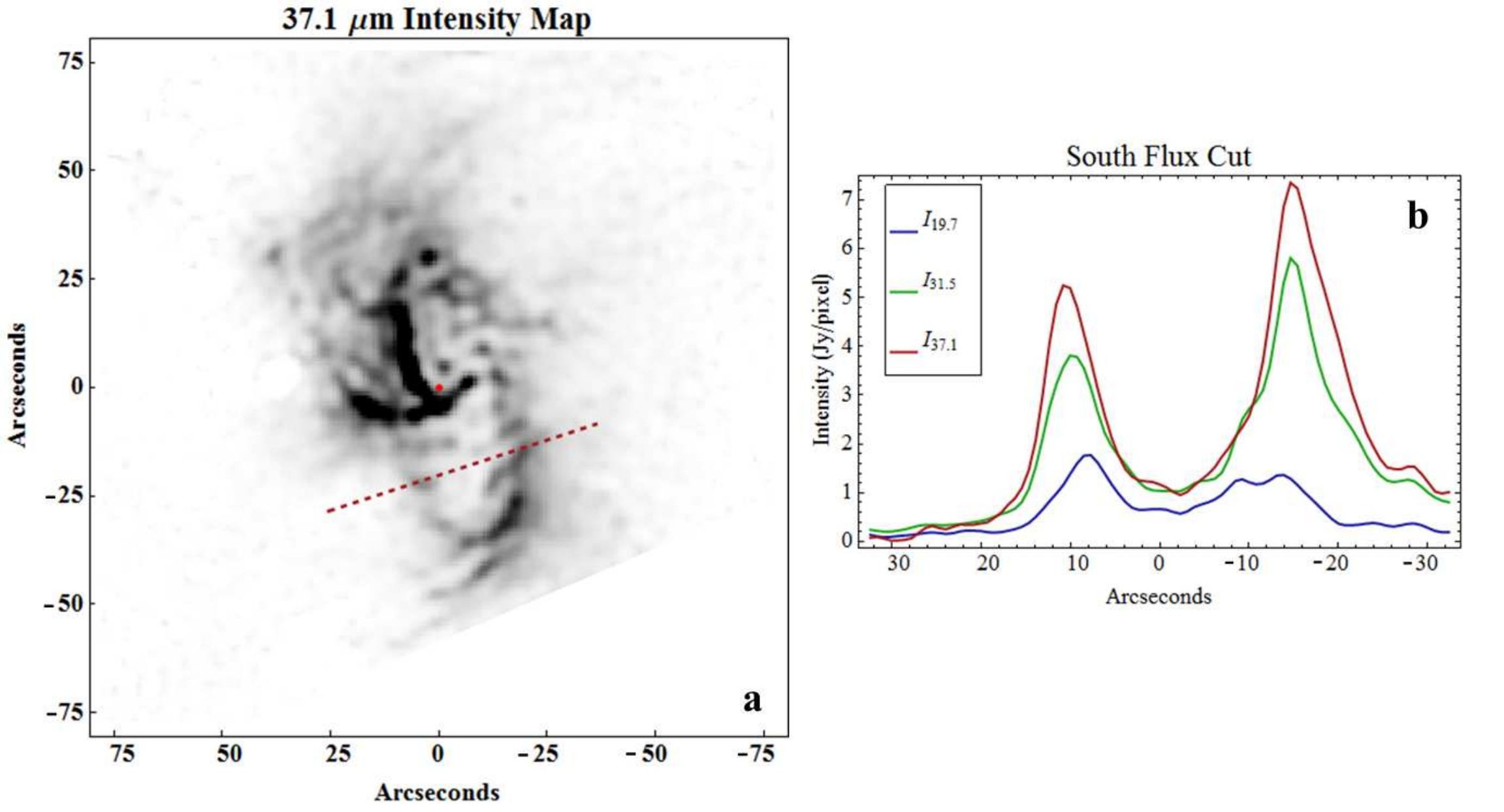}}
	\caption{(a) Deconvolved $37.1$ $\mu$m intensity map of the inner 6 pc of the GC overlaid with lines along which the intensities of the $37.1$, $31.5$, and $19.7$ $\mu$m maps are extracted. The red dot indicates the location of Sgr A*. (b) Intensities from the $37.1$, $31.5$, and $19.7$ $\mu$m maps along the southern line cut shown in (a) centered on (7'', -24''). The intensity plots shows noticeable shifts of the peak locations across the wavebands at the east and west regions of the CNR indicating that the observed dust in the CNR is heated centrally.}
	\label{fig:forcuts}
\end{figure}

We observe a distinct shift in the location of the peak intensities amongst the 19.7, 31.5, and 37.1 $\mu$m intensity maps. The 19.7 $\mu$m intensity peak is displaced radially inward from the 31.5 and 37.1 $\mu$m intensity peaks by several arcseconds, which indicates that the heating sources of the CNR are located in its interior. An intensity line cut through the southern region of the CNR is shown in Fig.~\ref{fig:forcuts}, illustrating the peak shifts. This observation strongly suggests that the CNR is identical to a classic HII region where the 19.7 $\mu$m dust emission primarily traces the ionized gas region while the 31.5 and 37.1 $\mu$m emissions extend into the photo-dissociation region (c. f. Salgado et al. 2012). Our observations are consistent with the claim that the Western arc is the ionized inner edge of the CND excited by UV radiation from the central stellar cluster (Serabyn \& Lacy 1985; Telesco et al. 1996). 

\begin{figure}[h]
	\centerline{\includegraphics[scale=.5]{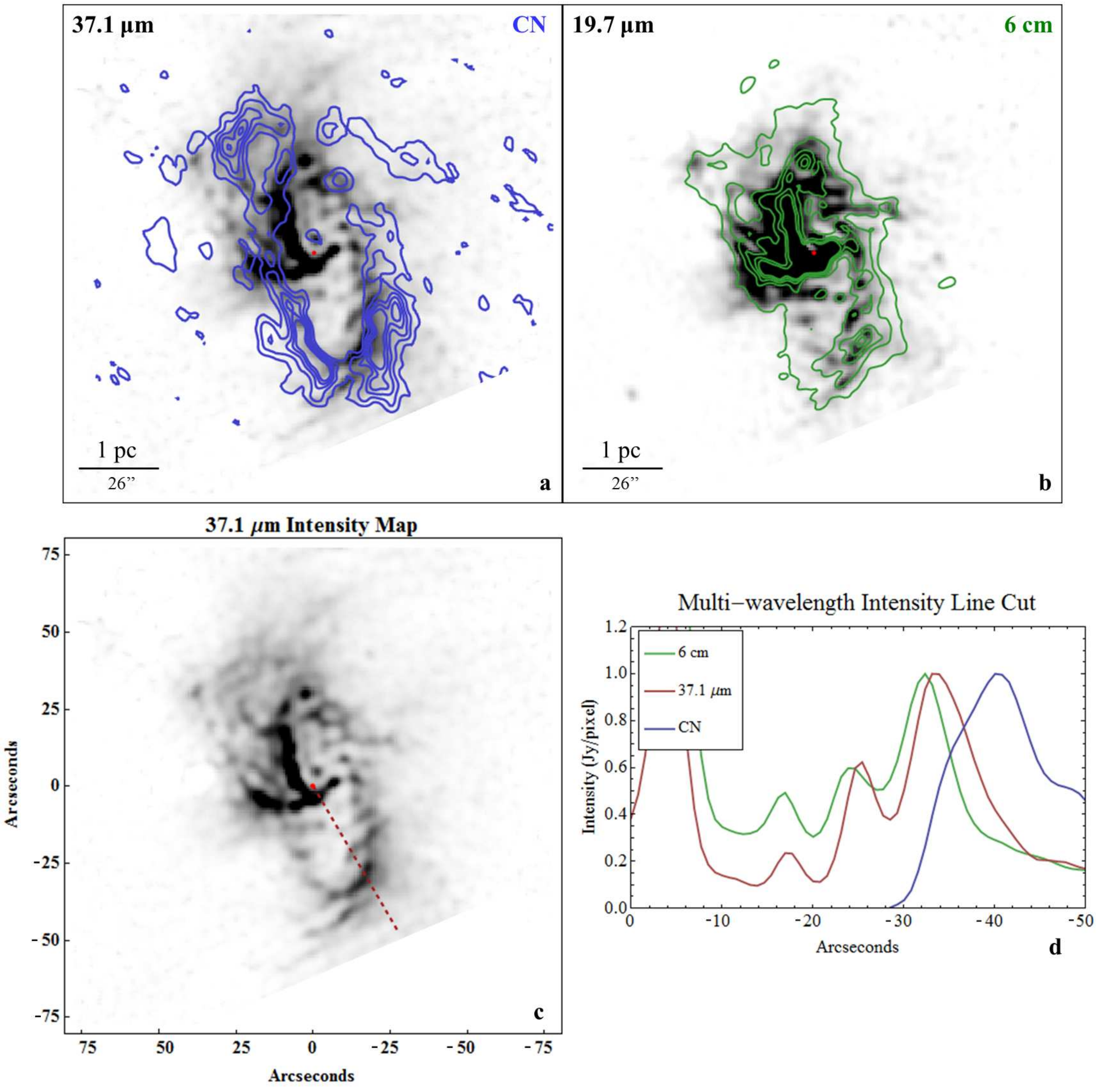}}
	\caption{(a) Intensity map at 37.1 $\mu$m overlaid with CN $2-1$ emission contours (Mart\'in et al. 2012). The lowest CN contour level corresponds to $10\sigma$ with steps increasing by $15\sigma$ ($\sigma=2.3$ Jy/Beam); the CN beam size has a FWHM of 4'' $\times$ 3''. (b) Intensity map at 19.7 $\mu$m overlaid with 6 cm radio emission contours (Yusef-Zadeh \& Morris et al. 1987) with levels corresponding to 1.0, 1.7, 2.4, 3.2 and 3.9 mJy/beam; the 6 cm beam size has been convolved to the same 2.5'' FWHM Gaussian as our images. The CN emission traces the cooler material in the CNR found beyond the 37.1 $\mu$m emission region while the 6 cm emission traces the ionized gas coincident with the 19.7 $\mu$m emission. (c) 37.1 $\mu$m intensity map overlaid with a line cut intersecting the southern region of the CNR. The intensity plot in (d) shows the fluxes across the 3 wavelengths along the cut overlaid in (c) and is normalized to the peak intensity of the CNR along the cut. The outwards progression of HII region (6 cm), PDR (37.1 $\mu$m), and molecular region (CN) indicates that the CNR is being heated by a centrally located source.}
	\label{fig:CNR6cmCN}
\end{figure}

In Fig.~\ref{fig:CNR6cmCN}a and b we overlay CN $2-1$ (Mart\'in et al. 2012) and 6 cm radio continuum (Yusef-Zadeh \& Morris 1987) contours on the 37.1 and 19.7 $\mu$m intensity maps, respectively. The CN contours trace the cooler material beyond the 37.1 $\mu$m emission region around most of the CNR while the radio contours are coincident counterparts to the 19.7 $\mu$m emission region. Fig.~\ref{fig:CNR6cmCN}d shows the normalized intensity line cut through the southern CNR (Fig.~\ref{fig:CNR6cmCN}c) across the 6 cm, 37.1 $\mu$m, and CN maps and presents the displacement of these three different emission regions, which is consistent with central heating. The region of the ionized gas emission at 6 cm lies slightly radially inward from the 37.1 $\mu$m, photo-dissociation region emission. Similarly, the 37.1 $\mu$m emission region is displaced inwards relative to the molecular emission region. The morphological arrangement of these regions strongly reinforces the picture of the CNR as a centrally heated structure. 

\subsection{Observed Dust Properties: Temperature, Optical Depth, and Luminosity}

Large column densities of dust and gas lead to extreme extinction along lines of sight towards the Galactic Center ($A_V \sim 30$) (Cardelli et al 1989). The extinction towards the Galactic Center has been characterized by numerous extinction curves derived through various techniques (Cardelli et al. 1989; Rieke et al. 1989; Lutz et al. 1999; Fritz et al. 2011). One of the primary sources of extinction is silicate dust grains, which absorb strongly at $9.7$ and $19$ $\mu$m. In this paper we adopt the curve of Fritz et al. (2011, hereafter referenced as F2011) derived from hydrogen emission lines of the minispiral from 1 - 19 $\mu$m observed by Short Wave Spectrometer (SWS) on the Infrared Space Observatory (ISO) and the Spectrograph for Integral Field Observations in the Near Infrared (SINFONI) on the Very Large Telescope (VLT). We select the F2011 extinction curve over the others because F2011 utilizes the most recent NIR (1 - 2.4 $\mu$m) observations of the Galactic Center. The F2011 curve is characterized by a -2.11 power-law slope at wavelengths shortward of 2.8 $\mu$m, absorption peaks in the mid-IR due to composite grains (in addition to silicates and carbonaceous grains), a 9.7 $\mu$m silicate feature optical depth of $\Delta \tau_{sil}=3.86$, and a $K_S$-band extinction of 2.62. Longwards of 24 $\mu$m we extrapolate from the F2011 extinction curve by adopting the Draine (2003) interstellar extinction curve defined in Fig. 10 of that paper. The extinction values at 24 microns agree between the F2011 and Draine (2003) extinction curves. From the extrapolated F2011 extinction curve, we therefore find that the ratio of the extinction at 19.7 to 31.5 $\mu$m, $e^{\tau_{19.7}}/e^{\tau_{31.5}}$, and at 19.7 to 37.1 $\mu$m, $e^{\tau_{19.7}}/e^{\tau_{37.1}}$, is 1.94 and 2.36, respectively.

After properly dereddening our images by applying the extrapolated F2011 extinction curve, we derive dust color temperature, optical depth, and total dust luminosity maps. We assume that the emission from the dust is optically thin and that the emissivity has a power-law form of $\nu^{\beta}$ and adopt an index of $\beta=2$, which is consistent with Draine (2003). The flux from the dust at $\nu$ can be expressed as $F_\nu \propto \nu^{\beta} B_\nu(T_D)$, where we define $T_D$ as the color temperature of the dust.

We use the derived color temperatures to estimate the emission optical depth, and with the power-law dust emissivity, the total dust luminosity is calculated as

\beq
L=4 \pi d^2\int_{\Omega_{\mathrm{source}}} \int_0^\infty F_{\nu_{37}}\left(\frac{\nu}{\nu_{37}}\right)^\beta \frac{B_{\nu} (T_D)}{B_{\nu_{37}} (T_D)}\,\mathrm{d}\nu\,\mathrm{d}\Omega,
\eeq

where $F_{\nu_{37}}$ is the observed flux at 37.1 $\mu$m, and $T_D$ is the color temperature determined from the observed 19.7/37.1 $\mu$m emission ratio.

\subsubsection{Color Temperature Map}

\begin{figure}[h]
	\centerline{\includegraphics[scale=.75]{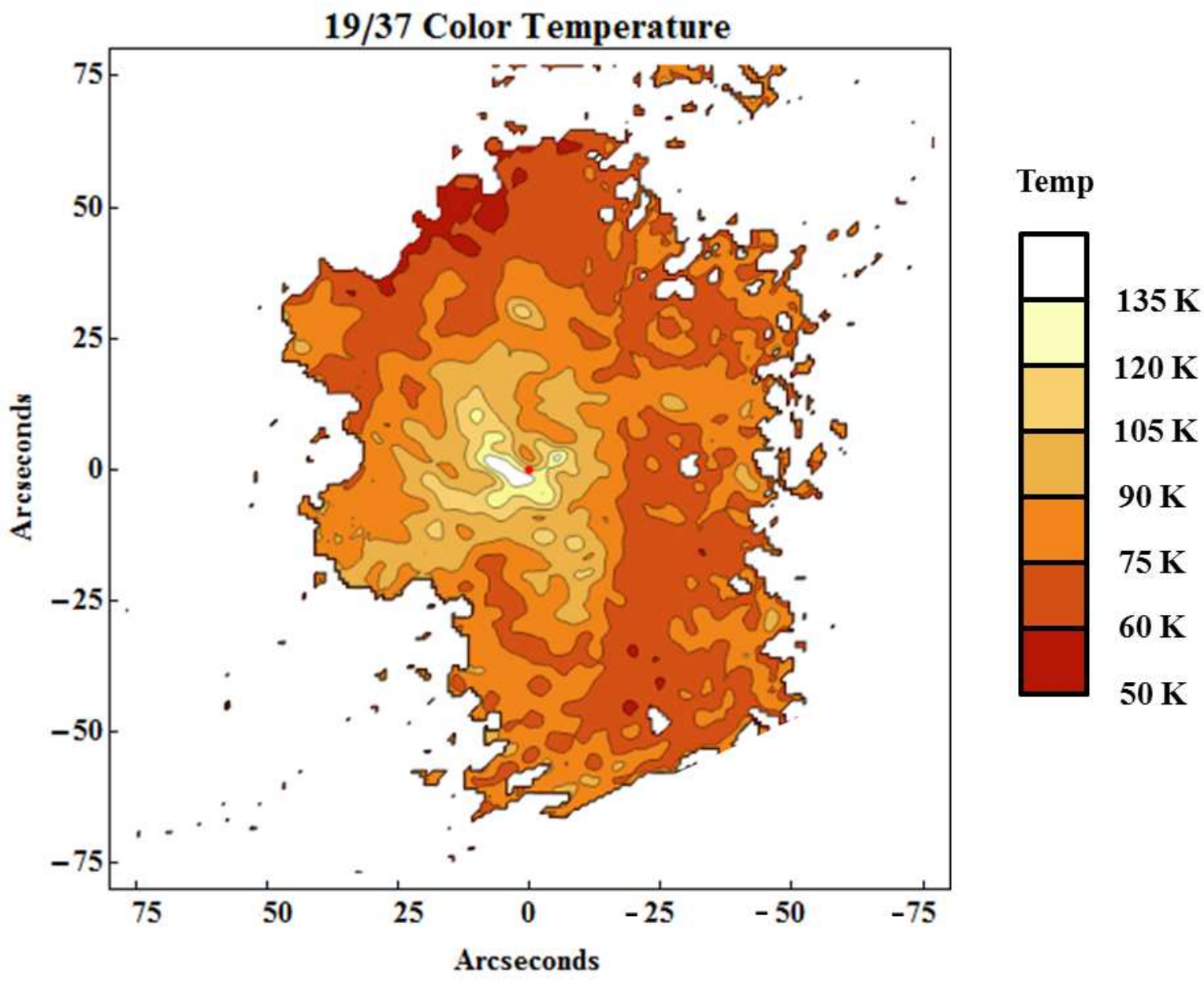}}
	\caption{19/37 $\mu$m color temperature contour map  of the inner 6 pc of the GC, calculated for pixels with signal-to-noise greater than 2$\sigma$. The contour levels correspond to temperatures of 50, 60, 75, 90, 105, 120, 135 (white) K. The red dot indicates the location of Sgr A*. Temperatures peak at 145 K in the region near Sgr A* where the Northern Arm appears to intersect with the East-West Bar. The CNR exhibits a decreasing radial temperature gradient with temperatures ranging from 65 - 85 K. Azimuthally uniform temperatures around the CNR as well as the decreasing radial temperature gradient  suggest heating of the CNR is dominated by central sources. The absence of sharp temperature fluctuations around the CNR indicates that no star formation is occurring within the ring.}
	\label{fig:CNRCT}
\end{figure}

The 19.7/37.1 $\mu$m color temperature map of the inner 6 pc of the Galactic Center is shown in Fig.~\ref{fig:CNRCT}. Given an absolute photometric error of $20\%$ the temperatures we derive are accurate within $\pm 8$ K (ignoring uncertainties in $\beta$). The highest temperature in the field peaks at 145 K at the Northern Arm $\sim8''$ to the east of Sgr A*. There is a decreasing radial temperature gradient going north along the Northern arm and going west along the East-West bar: temperatures in the Northern Arm range from 145 K to $\sim90$ K where it approaches the northern edge of the CNR, and the temperatures in the East-West bar range from 140 K to $\sim90$ K extending $\sim30$'' to the east of Sgr A* and past the eastern edge of the CNR. The CNR exhibits a radially outwards decreasing temperature gradient except for the eastern region, where the CNR dust emission is confused with that of the East-West Bar and Northern Arm. The temperatures range from 85 K at the inner edge of the ring to $\sim65$ K at the outer edge, consistent with central heating. Temperatures decrease radially due to the absorption of the dust-heating optical and UV photons. Other than the source IRS 8 ($\sim30$'' north of Sgr A*) we do not detect any embedded sources or large fluctuations in temperature, $\Delta T>15$ K, along the CNR. There is therefore no evidence from our observations of star formation occurring within the CNR.

\subsubsection{Optical Depth Map at 37.1 $\mu$m} 

\begin{figure}[h]
	\centerline{\includegraphics[scale=.75]{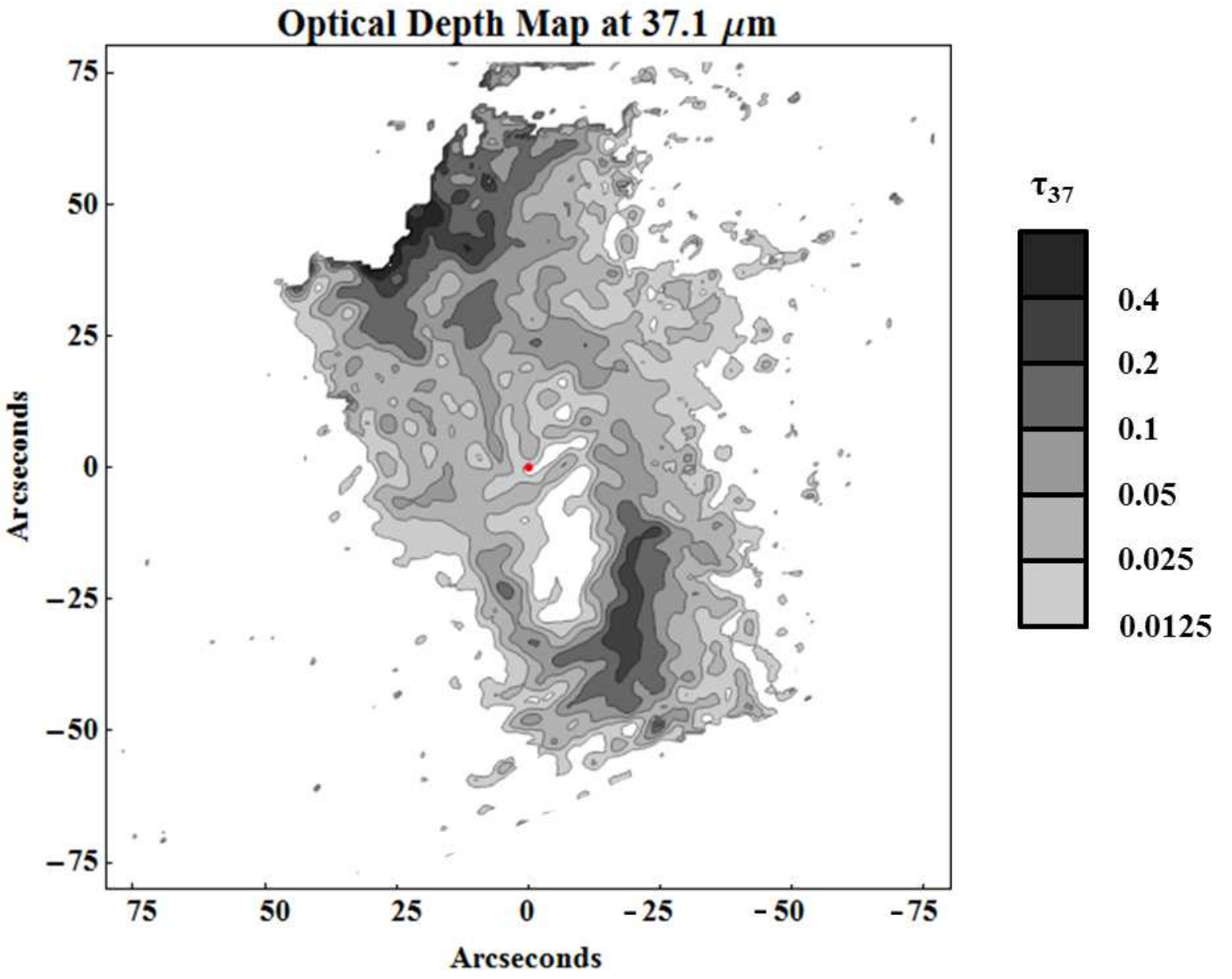}}
	\caption{Contour plot of the optical depth at 37.1 $\mu$m of the inner 6 pc of the GC derived from the 19/37 $\mu$m color temperature map and the 37.1 $\mu$m intensity map, calculated for pixels with signal-to-noise greater than 2$\sigma$. The contour levels correspond to $\tau_{37.1}$ of 0.0125, 0.025, 0.05, 0.1, 0.2, and 0.4 (black).  The red dot indicates the location of Sgr A*. Optical depths peak at 0.4 around the northern edge of the CNR. At the southern edge of the CNR the optical depths reach 0.3. The locations of the optical depth peaks are consistent with the effect of the high inclination ($67\,^{\circ}$) of the CNR increasing the column density along lines of sight at the northern and southern edges.  Much of the dust in the inner 6 pc of the GC lies around the CNR and not in the Northern Arm or East-West Bar. }
	\label{fig:CNROD}
\end{figure}

\begin{figure}[htbp]
	\centerline{\includegraphics[scale=.75]{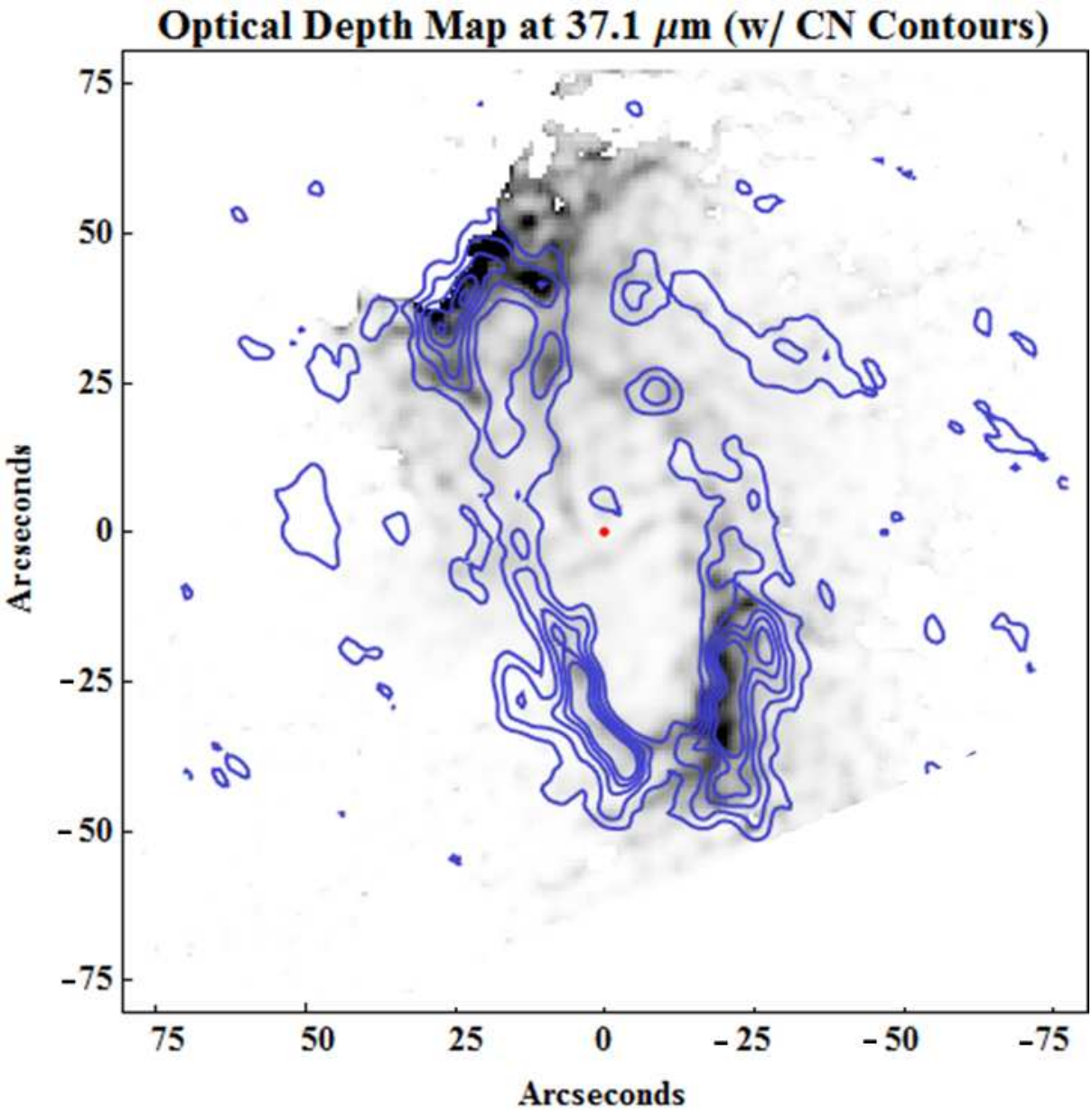}}
	\caption{Optical depth map at 37.1 $\mu$m overlaid with CN contours (Mart\'in et al. 2012). The contour levels are the same as in Fig.~\ref{fig:CNR6cmCN}a and the optical depth map is calculated for pixels with signal-to-noise greater than 2$\sigma$. The CN contours closely trace the 37.1 $\mu$m optical depth.}
	\label{fig:CNRODCN}
\end{figure}

We derive the 37.1 $\mu$m optical depth map (Fig.~\ref{fig:CNROD}) from the 37.1 $\mu$m intensity map and the 19.7/37.1 color temperature map. The 37.1 $\mu$m optical depth peaks at the northern ($\tau_{37.1}\sim 0.4$) and southern ($\tau_{37.1}\sim0.3$) edges of the CNR and decreases towards the eastern and western edges along the ring to $\sim0.05$. The increased optical depths at the north and south regions of the CNR are consistent with the enhanced column densities at those locations resulting from projection effect accompanying the high inclination of the ring, given our geometric interpretation of the ring dimensions (Fig.~\ref{fig:CNRview}). Along the Northern Arm the optical depths range from 0.04 to 0.13, peaking 28'' to the northeast of Sgr A* where the Northern arm appears to intersect the northern CNR. The East-West Bar exhibits lower values of $\tau_{37.1}$, ranging from 0.03 to 0.06, and peaks at the apparent intersection with the eastern CNR. Although the intensities are greatest at the Northern Arm and East-West Bar, the majority of the emitting dust in the field ($\sim75\%$) is located in and around the CNR. We find that the 37.1 $\mu$m optical depth traces the inner edge of the molecular CN emission (Mart\'in et al. 2012) along the Western Arc (Fig.~\ref{fig:CNRODCN}). At the northern region of the CNR the 37.1 $\mu$m optical depth appears to closely trace the CN emission; we suggest this may be due to the shadowing caused by the Northern Arm and East-West Bar which permits CN production to occur closer to the inner edge of the CNR than in the southwest.

The total gas mass of the CNR is derived and discussed in Sec. 4.3.

\subsubsection{Luminosity Map}

\begin{figure}[h]
	\centerline{\includegraphics[scale=.75]{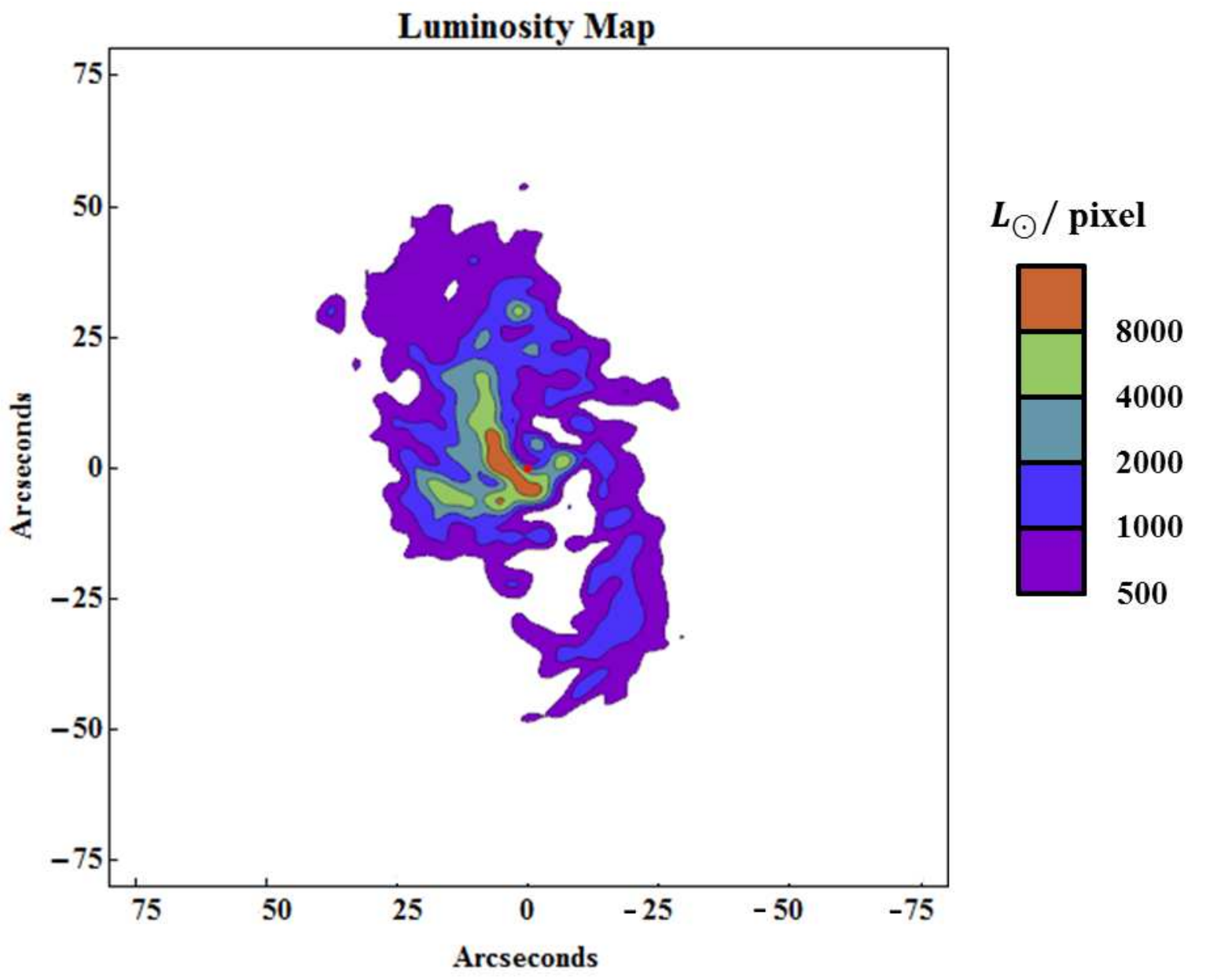}}
	\caption{Integrated luminosity contours derived from the 19/37 $\mu$m color temperature map and the 37.1 $\mu$m intensity map, calculated for pixels with signal-to-noise greater than 2$\sigma$. The contour levels correspond to 500, 1000, 2000, 4000, 8000 $\mathrm{L}_\odot$/pixel. The red dot indicates the location of Sgr A*. The luminosity contours closely trace the 37.1 $\mu$m intensity map. The total luminosity integrated over the inner 6 pc region is $\sim9.7 \times 10^6\,\mathrm{L}_\odot$.}
	\label{fig:CNRlum}
\end{figure}

In Fig.~\ref{fig:CNRlum} we present the integrated luminosity contours derived from the 19.7/37.1 color temperature map and the 37.1 $\mu$m map. We find that the luminosity contours closely trace the 37.1 $\mu$m intensity map with the only difference being that the Northern Arm and East-West Bar appear much more prominent in the luminosity map relative to the CNR since the temperatures are much greater along those features. The luminosities of the Northern Arm and East-West Bar are $\sim1.7\,\times\,10^6\,\mathrm{L}_\odot$ and $\sim1.8\,\times\,10^6\,\mathrm{L}_\odot$, respectively. Since the Northern Arm and East-West Bar may be shadowing the eastern CNR (Latvakoski et al. 1999) we estimate the expected, total luminosity of the CNR by multiplying the Western Arc luminosity by a factor of two. We derive an expected, total CNR luminosity of $\sim2.5\,\times\,10^6\,\mathrm{L}_\odot$.

There is a ``diffuse" component of the luminosity map not associated with the CNR, Northern Arm, and East-West Bar that has a luminosity of $\sim2.0\,\times\,10^6\,\mathrm{L}_\odot$ integrated over the entire field of view. The total luminosity integrated over the inner 6 pc region is $\sim9.7\,\times\,10^6\,\mathrm{L}_\odot$. Luminosities of the various features are summarized in Table~\ref{tab:Lumdata}. 

The stellar cluster near Sgr A* contains hot O and B stars as well as evolved Wolf-Rayet stars (Krabbe et al 1995). The combined luminosity of the central cluster is estimated by Krabbe et al. (1995) to be $\sim2\,\times\,10^7\,\mathrm{L}_\odot$. Since the outer boundary of the CNR is determined by the extinction of optical and UV photons as opposed to being limited by the amount of dust present to be illuminated, we can assume that the dust in the CNR absorbs all of the optical and UV photons emitted by the central cluster over the solid angle subtended by the inner wall of the CNR (see Fig.~\ref{fig:CNRCross}). This allows us to perform a consistency check on the morphology derived by the observed intensity maps. The fraction of the CNR luminosity over the luminosity of the hot central cluster stars provides an estimate of the inner wall height, $h$, and the opening angle of the CNR, $\phi_0$;

\beq
\frac{L_\mathrm{CNR}}{L_\mathrm{Cent}}=\frac{2 \pi\,R\,h}{4\pi\,R^2},\,\,
\phi_0=2\,\mathrm{Arctan}\left(\frac{L_\mathrm{CNR}}{L_\mathrm{Cent}}\right).
\label{eq:coverangle}
\eeq

From the luminosity fraction we derive an inner wall height of $\sim0.35$ pc and an opening angle of $\sim14^{\circ}$, which are similar to the values derived from the observed width of the CNR in the intensity map. 

Inversely, the total central cluster luminosity can be derived from the observed CNR dust luminosity and opening angle. Given an opening angle of $14^{\circ}$ and a CNR luminosity of $\sim2.5\,\times\,10^6\,\mathrm{L}_\odot$ we determine the central luminosity to be approximately $1.6\,\times\,10^7\,\mathrm{L}_\odot$, which is consistent with other luminosities also derived by the dust emission and geometric analysis ($2\,\times\,10^7\,\mathrm{L}_\odot$ (Davidson et al. 1992); $2.3\,\times\,10^7\,\mathrm{L}_\odot$ (Latvakoski et al. 1999)).

\section{Discussion}
\subsection{Central Heating}
In order to check the consistency of the central heating argument we calculate the theoretical equilibrium temperature of the dust in the CNR heated by the O and B stars in the central cluster. We treat the cluster as a point source at the location of Sgr A* since the size of the cluster is small compared to the radius of the CNR. The dust temperature, $T_d$, can then be derived from the expression shown in Eq.~\ref{eq:Deq},

\beq
\pi  a^2Q_*(a)\, F_*=4 \pi  a^2\,Q_{\mathrm{dust}}(T_d,a)\,\sigma _{\mathrm{SB}}\,T_d{}^4,
\label{eq:Deq}
\eeq

where $Q_\mathrm{dust}$ is the dust emission efficiency averaged over the Planck function of dust of size $a$ and temperature $T_d$, $Q_*$ is the dust absorption efficiency averaged over the incident radiation field, and $F_*$ is the incident flux at the location of the dust grains. For silicate-type grains and a heating source with average temperature 35000 K (Lacy et al. 1980) we have $Q_\mathrm{dust}(T_d, a)=1.3\times10^{-6} \left(\frac{a}{0.1\,\mu\mathrm{m}}\right)\,T_d^2$ and $Q_*\approx1$ (Draine 2011). Assuming a distance of 1.4 pc, a total source luminosity of $\sim2\,\times\,10^7\,\mathrm{L}_\odot$ (Krabbe et al. 1995), and a uniform size distribution of silicate-type dust grains with a standard interstellar medium grain size of $\sim0.2$ $\mu$m we derive a theoretical equilibrium temperature of 90 K which is consistent with the color temperatures we observe at the inner edge of the CNR ($\sim85$ K).

Using the DUSTY radiative transfer code (Ivezic \& Elitzur 1997) to perform an additional check on the temperature calculation, we are able to find consistent grain size parameters, that give the observed color temperatures at the inner radius of the CNR. The DUSTY model parameters are summarized in Table~\ref{tab:Dustypar}.

We consider the possibility of shock-driven heating by the high-velocity winds from the central cluster. The total estimated power contributed by the kinetic energy of the winds with a velocity of 700 km/s and a mass-loss rate of $10^{-2}\,M_\odot/\mathrm{yr}$ (Geballe et al. 1984) is $4\times10^{5}\, L_\odot$, which is $\sim2\%$ of the stellar luminosity from the central cluster. We therefore conclude that the heating due to shocks is not significant compared to the radiative heating.

\subsection{Modeling the CNR Dust Emission}

\subsubsection{Defining the outer edge of the CNR}

Our picture of the CNR as the illuminated inner edge of a larger, cooler distribution of gas and dust (Fig.~\ref{fig:CNRview}) allows us to calculate how deeply UV and optical photons penetrate into the ring--the point where the optical depth is $\sim1$. We assume the dust properties determined by the DUSTY model shown in Table~\ref{tab:Dustypar} and an inner edge gas density of $n_{H_2}=1.0\,\times10^4\,\mathrm{cm}^{-3}$ (Davidson et al. 1992; Latvakoski et al. 1999). The distance to the $\tau_V=1$ point can be derived as:

\beq
\tau_V=1=\int_{R_\mathrm{in}}^{R_{\tau_V=1}}n(r) \,  \bar{\sigma}\,\mathrm{d}r,
\label{eq:tau}
\eeq

where $n(r)$ is the radial density power-law with an index of $-1$ (Davidson et al. 1992), $\bar{\sigma}$ is the total cross section integrated over the dust size distribution, and $r$ is the radial distance from Sgr A*. We calculate a value of $R_{\tau_V=1}=0.46$ pc which is similar to the observed width of the CNR along the major axis ($\sim0.42$ pc) in the 37.1 $\mu$m intensity map. We note that changes in the radial density power-law index do not significantly alter the value of $R_{\tau_V=1}$; for an index of 1 we have $R_{\tau_V=1}=0.35$ pc.

\subsubsection{37.1 $\mu$m Intensity Model of the CNR}

We generate a 37.1 $\mu$m intensity model of the CNR assuming its morphology to be that of an inclined, circular ring (see Fig.~\ref{fig:CNRCross}). Instead of performing the full radiative transfer computations to determine the dust emission, we adopt a radial temperature power-law index of $-2/3$ which fits the observed temperature gradient along the major axis of the CNR in the 19/37 color temperature map. We note that a centrally heated, optically thin dust distribution would exhibit a radial temperature index of $-1/3$, assuming an emissivity power-law index of -2. The input parameters for the radius, $R$ (1.4 pc), inner edge temperature, $T_d$ (82 K), and inclination angle, $\theta_i$ ($67^{\circ}$), are taken from observations, and we are free to manipulate the inner edge density, $n_{H_2}$, and height (opening angle), $h$ ($\phi_0$), to fit the observed intensity map. The fitted model parameters are summarized in Table~\ref{tab:CNRmod}. Although the emitting region of the CNR is small compared to its radius, we must treat the geometry of the emitting region in careful detail since the opening angle, $\phi_0$, and the illumination depth, $R_{\tau_V=1}$, will largely determine the shape of the emission peak. 

\begin{figure}[h]
	\centerline{\includegraphics[scale=.4]{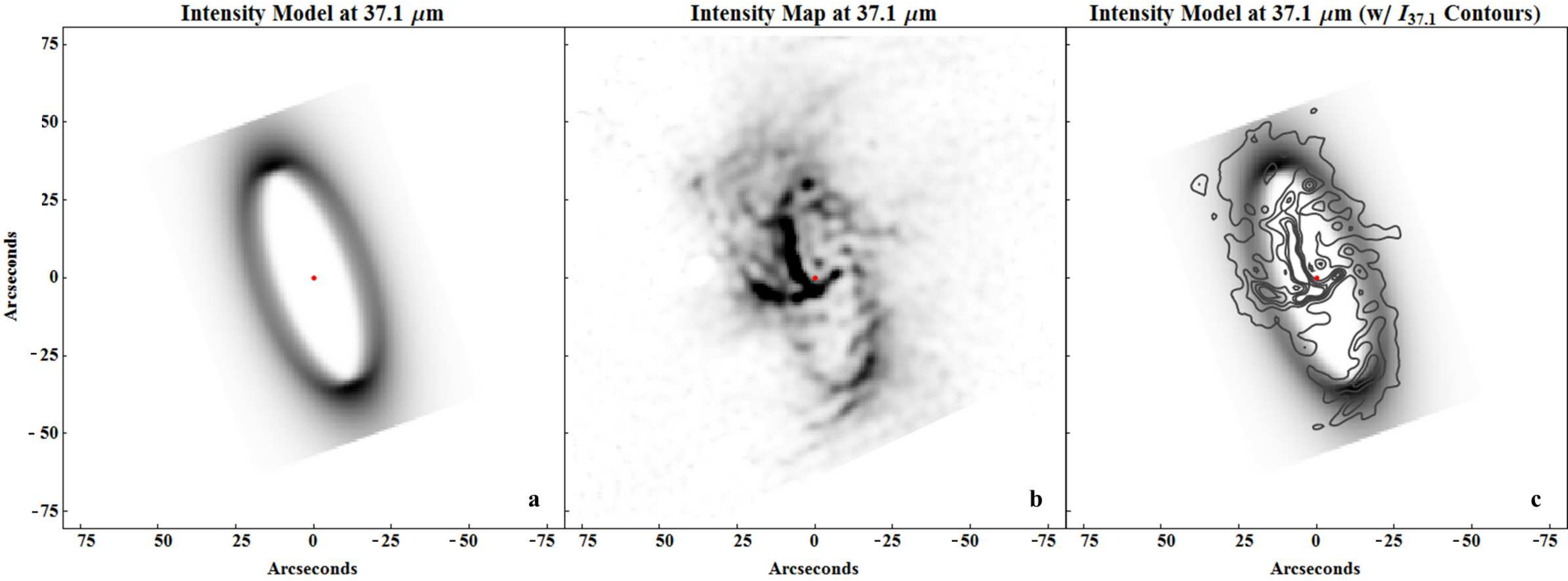}}
	\caption{(a) 37.1 $\mu$m CNR intensity model on the same scale as the 37.1 $\mu$m intensity map shown in (b). (c) 37.1 $\mu$m intensity contours overlaid on the CNR intensity model. The intensity contours correspond to 3, 6, 9, 12, and 15 Jy/pixel. The red dot indicates the location of Sgr A*}
	\label{fig:CNRMod}
\end{figure}

The intensity model is shown in Fig.~\ref{fig:CNRMod}a on the same size scale as the  intensity map in Fig.~\ref{fig:CNRMod}b and has been convolved with a 2.5'' FWHM Gaussian to simulate the beamsize of the deconvolved images. Contours of the 37.1 $\mu$m intensity map are overlaid on the model in Fig.~\ref{fig:CNRMod}c. Intensities in our model range from 5 Jy/pixel along the minor axis to 10 Jy/pixel along the major axis, which is consistent with the average intensities along the minor and major axes in the observed 37.1 $\mu$m intensity map. The observed intensity contours closely trace the model; however, the southern emission peak in the Western Arc is shifted several arcseconds to the north-west of the corresponding model peak. 

\begin{figure}[h]
	\centerline{\includegraphics[scale=.5]{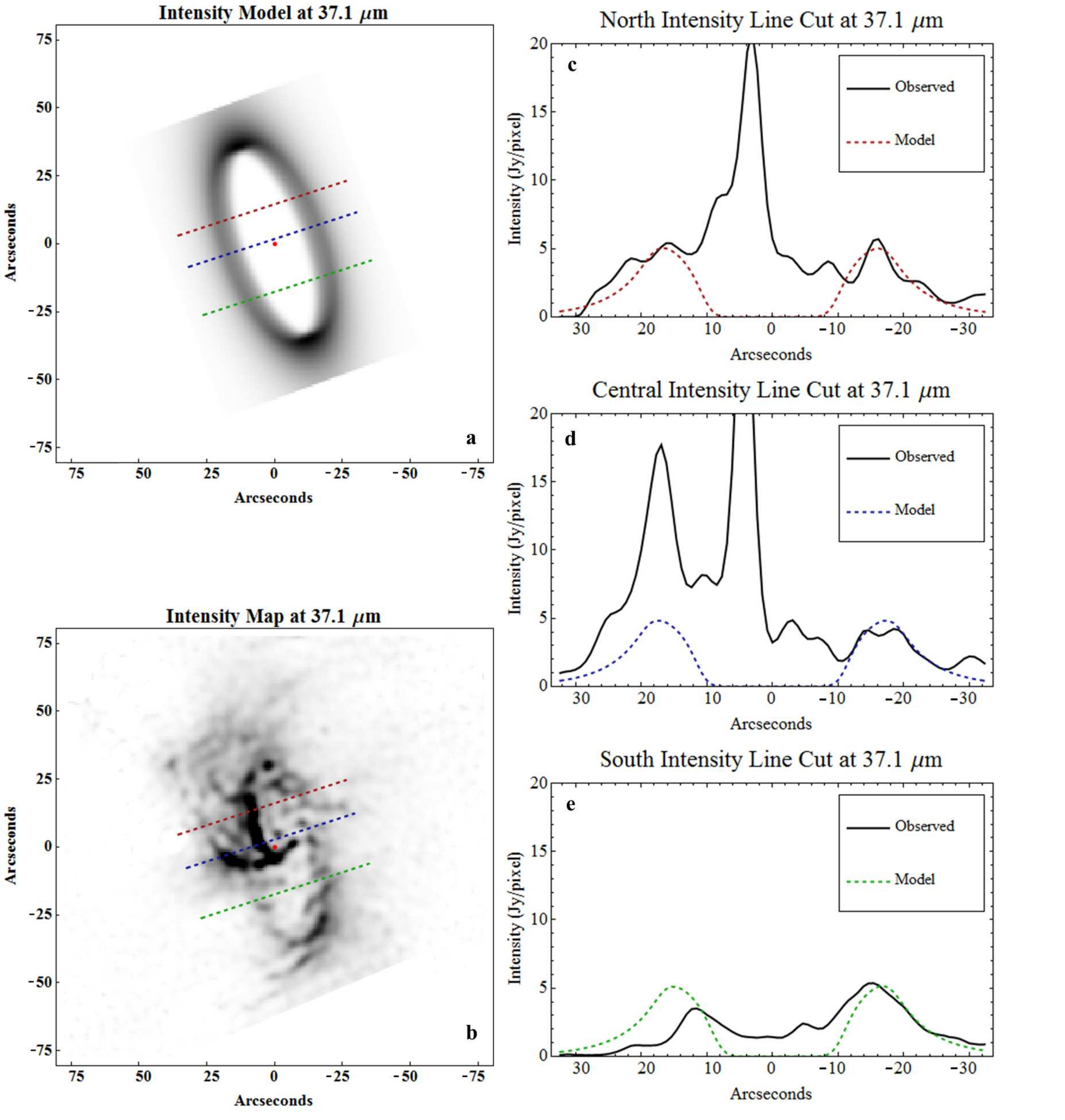}}
	\caption{(a) 37.1 $\mu$m CNR intensity model overlaid with line cuts perpendicular to the CNR along which the intensities are extracted. (b) 37.1 $\mu$m intensity map overlaid with the same line cuts as (a). (c) Northern intensity line cut centered on the point (-6'', 19''). (d) Central intensity line cut centered on the point (-2'', 3''). (e) Southern intensity line cut centered on the point (5'', -21''). The intensity, location, and shape of the western model peaks closely follow that of the observed peaks at the Western Arc. The peak near the origin of the intensity profiles in (c) and (d) corresponds to the emission from the Northern Arm. The intensity profiles in (e) show the difference ($\sim60\%$) in the emission from the western and eastern regions of the CNR. (Note the line cuts are intentionally placed to avoid intersecting the ``clumps" at the inside edge of the CNR)}
	\label{fig:CNRModObsCuts}
\end{figure}

A series of intensity line cuts across the model and observed CNR is shown in Fig.~\ref{fig:CNRModObsCuts}. The line cuts were positioned to avoid intersecting the ``clumps" at the inner edge of the CNR which is discussed in the following section. The location and amplitude of the model peaks at the Western Arc agree with the peaks from the observed intensity map. In the central (Fig.~\ref{fig:CNRModObsCuts}d) and southern (Fig.~\ref{fig:CNRModObsCuts}e) intensity line cuts even the widths of the model peaks closely agree with the observed intensity peak widths. The southern intensity line cut is the only cut that intersects with the observed eastern CNR; the emission at the eastern peak is $\sim60\%$ of the western peak hypothetically because the Northern Arm and East-West Bar absorb some of the UV and optical photons from the central cluster (Latvakoski et al. 1999). We do not observe the eastern CNR peak in the north and central line cuts due to confusion with the emission from the Northern Arm and East-West Bar. 

\subsection{CNR Dust Mass and Density}

We derive the total dust mass of the CNR using the 37.1 $\mu$m optical depth map and the grain parameters from our 37.1 $\mu$m intensity model (Table~\ref{tab:Dustypar}) and assuming a gas-to-dust mass ratio of 100. Integrating over the ring in the 37.1 $\mu$m optical depth map we find a total CNR mass of $\sim610\,M_\odot$, which is roughly $\sim1/2$ of the mass derived by Latvakoski et al. (1999) who observed the CNR at 31 and 37 $\mu$m. The discrepancy with Latvakoski et al. (1999) is due to assuming different grain parameters. Our derived mass is two orders of magnitude less than the mass derived by Etxaluze et al. (2011) and Requena-Torres et al. (2012), and roughly four orders of magnitude less than the mass derived by Christopher et al. (2005). This inconsistency can be attributed to difference in the tracers we used to study the CNR and, in the case of Christopher et al. (2005), the assumption of virialized clumps. Our observed dust emission does not trace the distribution of cooler material outside the CNR (deeper into the CND). 

The CNR gas density derived by our model ($n_{H_2}=10^4\,\mathrm{cm}^{-3}$) is consistent with the density derived observationally from our 37.1 $\mu$m optical depth map. Assuming the morphological parameters of our model the apparent thickness along the line of sight through the major axis of the CNR is $\sim1$ pc with a corresponding 37.1 $\mu$m optical depth of $\sim0.3$, which implies a density of $\sim1.4\times\,10^4$. The gas densities we find are similar to those of other works studying the CNR dust emission (Davidson et al. 1992; Latvakoski et al. 1999). 

\subsection{Differential Extinction Along the Western Arc}

In our analysis we assumed that the total extinction towards the CNR is constant around the ring; however, there is local extinction from the molecular component of the CND which, whose effect along the line of sight varies around the ring due to its inclined geometry. Christopher et al. (2005) suggest that the dust emission from the CNR does not trace all of the dust in the region due to extinction from high-density material in the molecular region. In this section, we therefore check for significant variations in our results when addressing the local, differential extinction.

\begin{figure}[h]
	\centerline{\includegraphics[scale=.5]{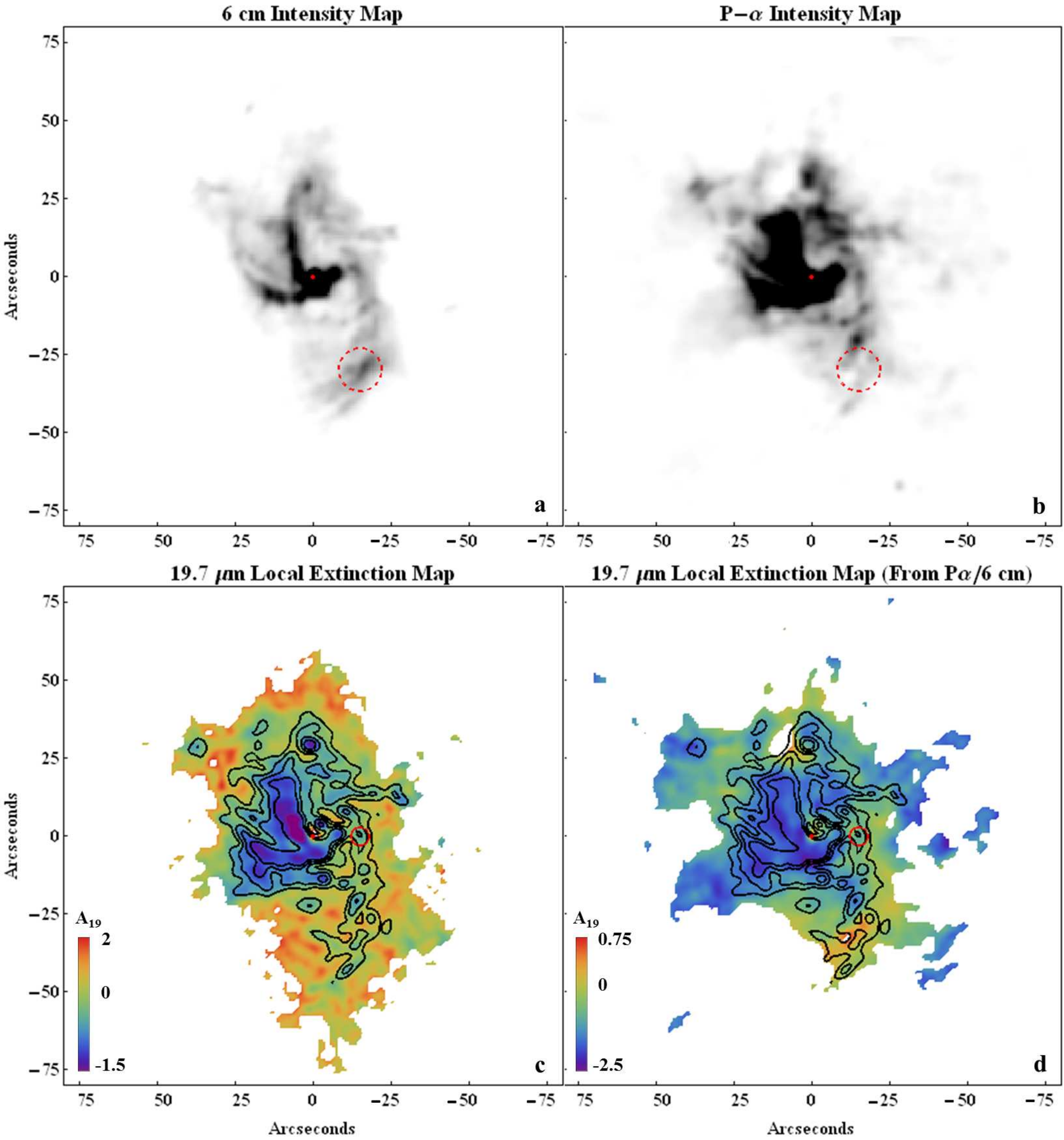}}
	\caption{(a) 6 cm and (b) continuum-subtracted Paschen-$\alpha$ (1.87 $\mu$m) intensity maps convolved to a 2.5'' FWHM Gaussian PSF. The Paschen-$\alpha$ emission at the circled region (-15'', -30'') decreases significantly relative to the intensity from the central Western Arc and indicates the presence of local, differential extinction. (c) 19.7 $\mu$m local extinction map derived from the 19 $\mu$m and 6 cm intensity maps. (d) 19.7 $\mu$m local extinction map derived from the Paschen-$\alpha$ and 6 cm intensity maps. Both (c) and (d) have the 19.7 $\mu$m intensity contours overlaid, with levels corresponding to 1, 2, 4, 8, and 16 Jy/pixel. The solid red circle (-15'', 0'') shows the reference region where the relative extinction was determined. The visual magnitude assumed for the interstellar extinction is $A_V=30$.}
	\label{fig:CNRExtComparison}
\end{figure}

 The effect of local, differential extinction can be seen by comparing the emission from the southern region of the Western Arc between the 6 cm and Paschen-$\alpha$ (Wang et al. 2010, Dong et al. 2011) intensity maps (Fig.~\ref{fig:CNRExtComparison}a and b). The emission at the southern Western Arc drops significantly between the 6 cm and the Paschen-$\alpha$ maps. Using the ratio of observed Paschen-$\alpha$ line flux to 6 cm radio free-free emission, we derive an absolute extinction map that agrees with that of Scoville et al. (2003). We find that the absolute extinction is greater along the Western Arc than toward the inner cavity and peaks at the southern portion of the CNR. After subtracting the interstellar extinction ($A_V=30$) we produce a map that represents only the local extinction (Fig.~\ref{fig:CNRExtComparison}d) towards the CNR. The local extinction along the Western Arc ranges from $A_{P\alpha}\sim1$ to 2 after correcting for interstellar extinction; this translates into a 19.7 $\mu$m (and 37.1 $\mu$m) extinction of less than one at the regions of peak extinction. Therefore, as a first approximation, the local, differential extinction derived by the Paschen-$\alpha$ and 6 cm maps is not significant at our wavelengths.

As mentioned in Sec. 3.2, the 19.7 $\mu$m emission primarily traces the ionized gas region of the CNR, which is where the Paschen-$\alpha$ and radio free-free emission originate. Unlike for Paschen-$\alpha$, we do not have a relation that provides us with the intrinsic 19.7 $\mu$m emission given the 6 cm emission. We therefore produce an extinction map relative to the extinction towards a fiducial region in the 19.7 $\mu$m and 6 cm maps (the red circle in Fig.~\ref{fig:CNRExtComparison}c and d). We  assume that the intrinsic, unextinguished emission at 19.7 $\mu$m and 6 cm are directly proportional at all locations around the ring: $I_{19\,\mu\mathrm{m}}^{\mathrm{Int}}\propto I_{6\,\mathrm{cm}}^{\mathrm{Int}}$. The relative extinction at 19.7 $\mu$m is $\tau_{19\,\mu\mathrm{m}}^{\mathrm{(x, y)}}-\tau_{19\,\mu\mathrm{m}}^{\mathrm{(ref)}}$, where $\tau_{19\,\mu\mathrm{m}}^{\mathrm{ref}}$ and $\tau_{19\,\mu\mathrm{m}}^{\mathrm{(x, y)}}$ are the extinction values towards the reference region and towards an arbitrary position on the map, respectively. Assuming the extinction at 6 cm is negligible, the relative extinction can be derived from the relation shown in Eq.~\ref{eq:relext} provides the relative extinction, where we have taken the extinction at 6 cm to be negligible.

\beq
\left(\frac{I_{19\,\mu\mathrm{m}}^{\mathrm{Obs}}}{I_{6\,\mathrm{cm}}^{\mathrm{Obs}}}\right)_{\mathrm{ref}}=\left(\frac{I_{19\,\mu\mathrm{m}}^{\mathrm{Int}}}{I_{6\,\mathrm{cm}}^{\mathrm{Int}}}\right)_{\mathrm{(x, y)}}e^{{\tau_{19\,\mu\mathrm{m}}^\mathrm{(x, y)}}-\tau_{19\,\mu\mathrm{m}}^{\mathrm{ref}}}
\label{eq:relext}
\eeq

We produce a map of the local extinction (Fig.~\ref{fig:CNRExtComparison}c) by referring to the local extinction map derived from the  Paschen-$\alpha$/6 cm ratio (Fig.~\ref{fig:CNRExtComparison}d) to determine the value of $\tau_{19\,\mu\mathrm{m}}^{\mathrm{ref}}$. The local extinction derived from the 19.7 $\mu$m/ 6 cm ratio ranges from $A_{P\alpha}\sim2.5$ to $\sim4$ (or $A_{19}\sim1$ to $\sim1.7$) along the Western Arc, which is somewhat larger than obtained using Paschen-$\alpha$ and 6 cm maps (see below). We apply this local extinction correction to the dereddened 19.7 and 37.1 $\mu$m images and find that the total flux along the Western Arc increases by $\sim60\%$ and $\sim15\%$, respectively. The corrected $19.7/37.1$ color temperature map does not exhibit deviations greater than $\pm 10$ K ($\sim15\%$) from the uncorrected color temperature map, which is on the order of our estimated error. Since the 37.1 $\mu$m flux and the color temperature do not change significantly we choose to ignore the effect of differential extinction and stand by our original analysis. Of course the assumptions of a constant 19.7 $\mu$m to 6 cm ratio likely introduces additional uncertainty to the intensity maps corrected this way.

As noted above, the local extinction derived from the 19.7 $\mu$m and 6 cm maps deviates from the extinction derived using the Paschen-$\alpha$ and 6 cm maps (Fig.~\ref{fig:CNRExtComparison}d) by a factor of $\sim2$. This suggests that the extinguishing material in the molecular region of the CND is very porous and ``clumpy" since the Paschen-$\alpha$ map will reveal only the emission unabsorbed by the intervening clumps, whereas the emission at 19.7 $\mu$m can penetrate through these same clumps and appear in the 19.7 $\mu$m map. The evidence of clumpiness is consistent with the analysis of the CND performed by Genzel et al. 1989 and Davidson et al. 1992.

\subsection{Characterizing the CNR ``Clumps"}
\label{sec:clumps}

\begin{figure}[h]
	\centerline{\includegraphics[scale=.75]{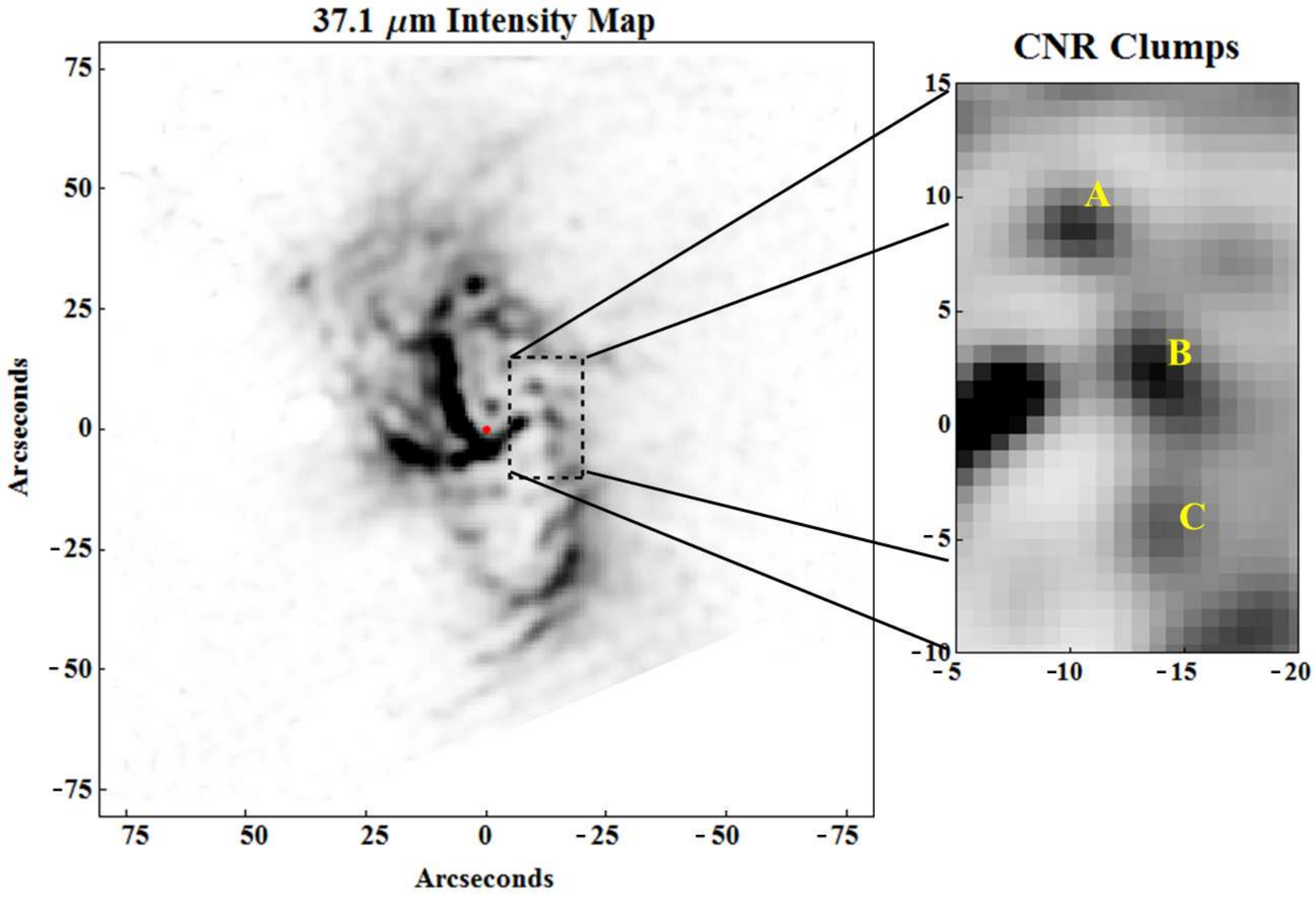}}
	\caption{37.1 $\mu$m intensity map of the inner 6 pc of the GC overlaid with the subregion containing three prominent clumps at the inner edge of the CNR labeled ``A'', ``B'', and ``C.''}
	\label{fig:CNRClumps}
\end{figure}

At the inner edge of the Western Arc of the CNR (Fig~\ref{fig:CNRClumps}) we observe several small ``clumps" that have a FWHM of $\sim 4-5''$ ($\sim0.15$ pc). The clumps are regions of enhanced density relative to the CNR as opposed to being temperature peaks heated by an embedded source. The clumps are displaced radially inward from the CNR; however, radial velocity observations indicate that the clumps are indeed part of the CNR (Irons et al. 2012). We identify three clumps along the central, inner edge of other CNR labeled ``A", ``B", and ``C" from north to south.  Clump B has a strong radio counterpart (Yusef-Zadeh and Morris 1987) while clump A has a prominent molecular counterpart observed in HCN emission (Christopher et al. 2005). We derive the masses and densities of the clumps from the 37.1 $\mu$m optical depth map and assume their volume can be estimated by $\sim\frac{4}{3}\,\pi \, (\mathrm{FWHM}/2)^3$. The clump properties are summarized in Table~\ref{tab:Clumppar}. We find that the clumps exhibit densities ranging from 5 to $9\,\times\,10^4\,\mathrm{cm}^{-3}$ which are consistent with the low-excitation clump densities derived by Requena-Torres et al. (2012) in their large velocity gradient analysis of CO observations at the northern and southern regions of the CND. Requena-Torres et al. (2012) derive slightly larger clump sizes ($r\sim0.3$ pc) than we do but their estimate is not well constrained since it is partly degenerate with the kinetic temperature. Under the assumption that the clumps are virialized, Christopher et al. (2005) determine a mass and density for clump A (core V in Christopher et al. 2005) of $\sim2400\,M_\odot$ and $n\sim10^7\,\mathrm{cm}^{-3}$ from the HCN emission.The virial density estimate would imply that the clumps are stable against tidal disruptions; however, our density estimates, as well as those of Requena-Torres et al. (2012), suggest the clump densities are a factor of $\sim10^3$ too low to meet the stability requirement. 

We perform a simple calculation to estimate the lifetime of the clumps assuming that they are not virialized, so that their lifetime is limited by the rate at which they are azimuthally sheared as the CNR undergoes differential, Keplerian rotation around Sgr A*. Given the approximate extent of the clumps in the radial direction from Sgr A* ($\sim0.15$ pc), we identify the recognizable lifetime of a clump with the time required for its opposite sides to separate 
azimuthally by a distance equal to the azimuthal separation of the clumps, $\sim0.3$ pc. This yields a clump lifetime of $\sim40000$ years, or about half of the orbital period of the CNR.

\begin{figure}[h]
	\centerline{\includegraphics[scale=.75]{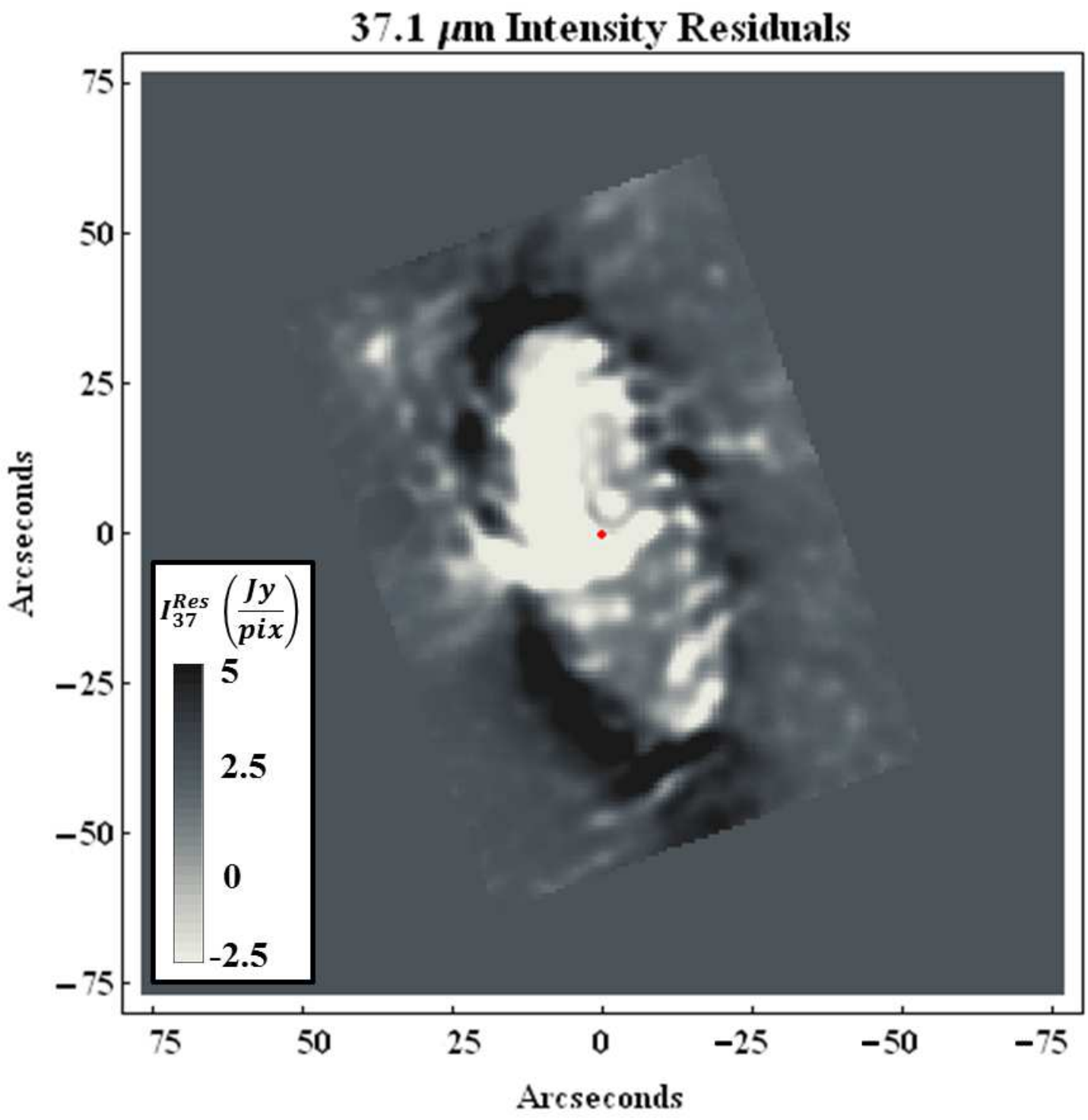}}
	\caption{Intensity residuals from subtracting the 37.1 $\mu$m intensity model from the 37.1 $\mu$m map. Light and dark regions correspond to positive and negative intensity residuals, respectively. The intensity model is based on the Western Arc and hence oversubtracts from the shadowed eastern and northern sections of the ring. Along the inner edge of the CNR are the clumps which exhibit excess emission compared to the CNR. Behind almost every clump is a region of negative intensity indicating that the clumps shadow the CNR material behind them whereas between the clumps in the CNR the residual intensities are close to zero.}
	\label{fig:IntensityRes}
\end{figure}

Fig.~\ref{fig:IntensityRes} shows the observed 37.1 $\mu$m intensity residuals after subtracting the 37.1 $\mu$m intensity model (Fig.~\ref{fig:CNRMod}). The clumps appear prominently at the inner edge of the Western Arc where the emission is greater than that predicted by the model, while inbetween the clumps along the ring the residuals are near zero. Interestingly, we find that the intensity model over-predicts the emission ``behind'' the clumps implying that they are likely shadowing the material further within the CNR. Given our derived dust grain parameters and assuming a clump density of $5\times 10^{4}\,\mathrm{cm}^{-3}$ we calculate that the distance to the point at which $\tau_\mathrm{V} = 1$ within the clump is $\sim0.08$ pc which is on the order of the projected clump width; therefore, the clumps may indeed be shadowing the CNR material. 

While the CND is not a fully relaxed system, its relatively high degree of order and symmetry (e.g., Jackson et al. 1993) suggest that it has undergone at least a few rotations since its formation. Consequently, the very existence of such clumps is difficult to understand if they are not self-gravitating. We will turn to this key question in a subsequent paper. 

\section{Conclusions}
We have presented images of the inner 6 pc of the Galaxy with a spatial resolution of 3.2 - 4.6'' (2.5'' deconvolved) at 19.7, 31.5. and 37.1 $\mu$m. From the images we determine the morphology and structure of the CNR to be that of an inclined ($67^{\circ}$) 1.4 pc radius ring centered on Sgr A*. The CNR is thin relative to its radius with an inner wall height of $\sim3$ pc. Our morphological interpretation agrees with the previous 31 and 37 $\mu$m dust emission analysis done by Latvakoski et al. (1999). The progressively outwards physical locations of the 6 cm, 37.1 $\mu$m, and CN emission regions show that the CNR is ionized and heated by central sources. Our temperature maps verify this picture and reveal a slightly decreasing radial temperature gradient with an inner ring temperature (85 K) consistent with theoretical calculations assuming a central point source of $2\times10^7\,\mathrm{L}_\odot$ with a temperature of $35000$ K. We generate a 37.1 $\mu$m intensity model of the CNR by adopting the observed temperatures and morphological properties. We find that our intensity model agrees with the observed 37.1 $\mu$m emission from the Western Arc.

We have addressed the issue of local, differential extinction along the ring by comparing both the Paschen-$\alpha$ and 19.7 $\mu$m intensity maps to the 6 cm radio map, and have concluded that it did not result in a significant change to the observed intensity at our wavelengths. We however discovered an inconsistency in the magnitude of extinction between the Paschen-$\alpha$ and 19.7 $\mu$m-derived extinction maps: the 19.7 $\mu$m-derived extinction values were a factor $\sim$ 2 to 3 times greater than the Paschen-$\alpha$ derived values when converted to the extinction at identical wavelengths. We resolved this discrepancy by invoking the clumpiness of the material in the molecular region of the CND.

Our optical depth maps of the CNR reveal that the clumpiness of the ring is due to density enhancements rather than embedded sources or sites of star formation. The density of the CNR between the clumps is $\sim 10^4 \,\mathrm{cm}^{-3}$, which agrees with previous density estimates derived from the dust emission (Davidson et al. 1992; Latvakoski et al. 1999). We find that the densities and sizes of the clumps along the inner edge of the CNR ($\sim$5 to 9 $\times 10^4\,\mathrm{cm}^{-3}$ and $\sim0.15$ pc) are consistent with the densities of low-excitation molecular clumps observed by Requena-Torres et al. (2012) at the southern and northern regions of the CNR. Clumps of this density and size will be sheared out by the tidal forces from the supermassive black hole on timescales of half an orbital period ($\sim40000$ yrs) at a distance of 1.4 pc from Sgr A*. The existence of these clumps at such a close distance to Sgr A* therefore presents an interesting problem which we will further investigate.

\emph{Acknowledgments}. We would like to thank the rest of the FORCAST team, George Gull, Justin Schoenwald, Chuck Henderson, and Jason Wang, the USRA Science and Mission Ops teams, and the entire SOFIA staff. We would also thank Hans Zinnecker and the referee for their valuable comments. This work is based on observations made with the NASA/DLR Stratospheric Observatory for Infrared Astronomy (SOFIA). SOFIA science mission operations are conducted jointly by the Universities Space Research Association, Inc. (USRA), under NASA contract NAS2-97001, and the Deutsches SOFIA Institut (DSI) under DLR contract 50 OK 0901. Financial support for FORCAST was provided by NASA through award 8500-98-014 issued by USRA.

\section{Figures}



\section{Tables}

\begin{deluxetable}{ccccccc}
\tablecaption{Sgr A West Luminosities and CNR Opening Angle}
\tablewidth{0pt}
\tablehead{$L_\mathrm{bkgd}$ ($L_\odot/\mathrm{pix}$) &$L_\mathrm{Tot}$&$L_\mathrm{Tot}-L_\mathrm{bkgd}$ &$L_\mathrm{NA}$ &$L_\mathrm{Bar}$&$L_\mathrm{CNR}$\tablenotemark{a} &$\phi_0$ ($^{\circ}$)\tablenotemark{b}}

\startdata
	$70\pm15$&9.8&$7.0$&1.7&1.8&2.5&$14\pm3$\\ 
\enddata

\tablecomments{Units given in $10^6\, L_\odot$}
\tablenotetext{a}{Expected, total CNR luminosity derived by multiplying Western Arc luminosity by a factor of two}
\tablenotetext{b}{Derived from Eq.~\ref{eq:coverangle}; total measured luminosity of observed stars in central cluster is $L_\mathrm{Cent}\sim2\times10^{7} \, L_\odot$ (Krabbe et al. 1995)}
	\label{tab:Lumdata}
\end{deluxetable}

\begin{deluxetable}{ccccccc}
\tablecaption{DUSTY Model Parameters.}
\tablewidth{0pt}
\tablehead{$T_d$ (K) & $R_\mathrm{in}$ (pc) & $L_\mathrm{Source}$ ($L_\odot$) & $a_\mathrm{min}$ ($\mu$m) & $a_\mathrm{max}$ ($\mu$m) & $\alpha$ & $m$}

\startdata
	82&1.4&$2\times10^7$&0.1&2& -3.5 & -1\\ 
\enddata

\tablecomments{$T_d$ is the dust temperature at the inner edge, $R_\mathrm{in}$ is the inner radius of the CNR, $L_\mathrm{Source}$ is the heating source luminosity, $a_\mathrm{min}$ is the minimum grain size, $a_\mathrm{max}$ is the maximum grain size, $\alpha$ is the power-law index of the grain size distribution, and $m$ is the power-law index of the radial density profile.}

	\label{tab:Dustypar}
\end{deluxetable}

\begin{deluxetable}{ccccccc}
\tablecaption{Observed and Model-Derived CNR Properties.}
\tablewidth{0pt}
\tablehead{ $R$ (pc) & $h$ (pc)&$\theta_i$&$T_d$ (K) & $q$& $n_{H_2}$ ($\times\,10^4\,\mathrm{cm}^{-3}$) & Mass from $\tau_{37}$\tablenotemark{a} ($M_\odot$)}

\startdata
	1.4&0.34&$67^{\circ}$&82&-2/3&1.0& $\sim610$\\ 
\enddata

\tablecomments{The radius, $R$, temperature, $T_d$, radial temperature index, $q$, inclination angle, $\theta_i$, and the mass are taken from observations. The CNR inner radius density, $n_{H_2}$, and height, $h$, are derived from the model.}

\tablenotetext{a}{Assuming a gas to dust mass ratio of 100}
	\label{tab:CNRmod}
\end{deluxetable}

\begin{deluxetable}{c||ccccc}
\tablecaption{Observed Clump Properties.}
\tablewidth{0pt}
\tablehead{Clump & $d$ (pc) & FWHM (pc)&$T_d$ (K) & $n_{H_2}$ ($\times\,10^4\,\mathrm{cm}^{-3}$) & Mass from $\tau_{37}$\tablenotemark{a}($M_\odot$)}

\startdata
	   A&1.14&0.13&$90$&5& $\sim$1.5  \\ 
           B&1.25&0.15&86&9& $\sim$2.0\\ 
           C&1.13&0.13&80&5& $\sim$2.5\\ 
\enddata

\tablenotetext{a}{Assuming a gas to dust mass ratio of 100}
	\label{tab:Clumppar}
\end{deluxetable}

\end{document}